\documentclass[a4paper,11pt]{article}
\usepackage{pos}
\usepackage{wrapfig}
\usepackage{makecell}
\usepackage{epstopdf}

\title{Aspects of finite temperature QCD towards the chiral limit}

\author*{Anirban Lahiri}

\affiliation{Fakult\"{a}t f\"{u}r Physik, \\ Universit\"{a}t Bielefeld.\\
  Bielefeld, Germany.}

\emailAdd{alahiri@physik.uni-bielefeld.de}

\abstract{QCD under extreme conditions has been studied for a long time, and the chiral limit has been a grey area mostly.
In this write-up of my talk, I review some of the recent developments made by the community to unveil various features of QCD
towards the chiral limit, which includes calculation of the chiral critical temperature and determination of the order
of chiral phase transition for various numbers of flavors. Acknowledging the importance of the studies regarding the
effective restoration of $U_A(1)$, I try to give a comprehensive overview about the various studies done in the last
few years in a comparative manner to realize the current status of the community in this regard. I also discuss
very recent efforts about the relevance of various energy-like observables w.r.t.\ the chiral phase transition.}

\FullConference{%
 The 38th International Symposium on Lattice Field Theory, LATTICE2021
  26th-30th July, 2021
  Zoom/Gather@Massachusetts Institute of Technology
}

\begin{document}
\maketitle

\section{Why this talk?}

Let me start with some prelude about why it is interesting to study
the Quantum Chromo-Dynamics (QCD) in/towards the chiral limit,
defined by the vanishing quark masses,
which can't be realized in nature.

Two fundamental features of QCD are spontaneous breaking of chiral symmetry and confinement. More than
four decades ago, there were studies to understand their
interplay. One set of studies claimed that the
spontaneous breaking of the chiral symmetry happens in a confining
environment \cite{Casher:1979vw,Banks:1979yr}. Almost at the same
time other studies tried to answer the question by calculating and
comparing the corresponding transition temperatures
\cite{Kogut:1982fn,Kogut:1982rt}, but this remains somewhat inconclusive
w.r.t.\ the behavior of theories with different number of colors.
Although not very sophisticated w.r.t.\ present day calculations,
the above-mentioned works shows the importance of the chiral
transition temperature, which in itself defines a fundamental scale.

\begin{wrapfigure}{r}{6cm}
\centering
\includegraphics[scale=0.4]{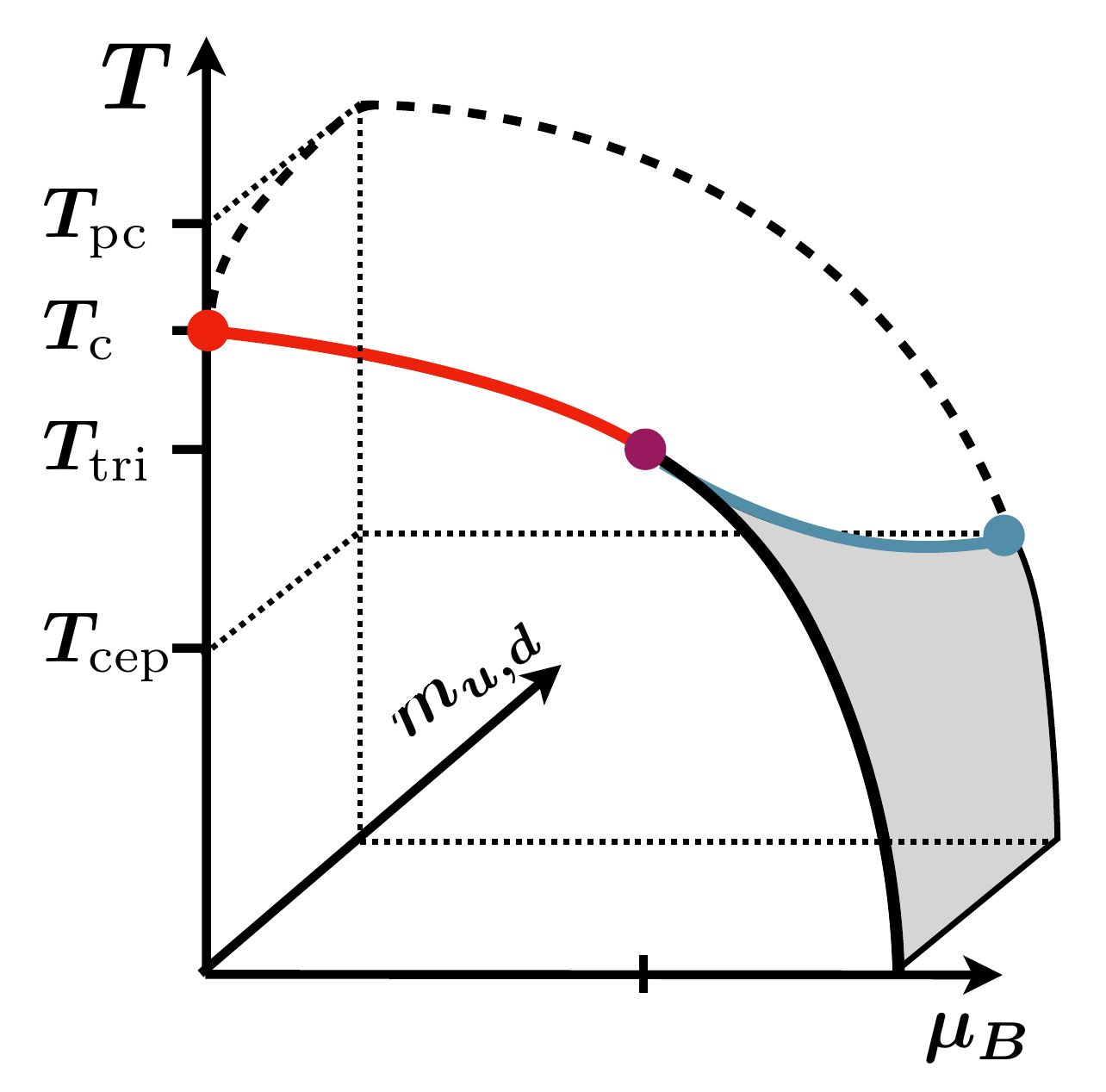}
\caption{Schematic phase diagram of QCD in the space of
temperature ($T$) $-$ baryon chemical potential ($\mu_B$) $-$
degenerate mass of $u$ and $d$ quarks. Figure is taken from
ref.\ \cite{Karsch:2019mbv}.}
\label{fig:3dQCDPD}
\end{wrapfigure}

On the phenomenological side also, the chiral transition temperature
and the nature of the chiral transition is very important to establish
the global phase structure of strongly interacting matter.
To illustrate this point, I show the conjectured phase diagram of
QCD in the space of temperature ($T$) $-$ baryon chemical potential
($\mu_B$) $-$ degenerate mass of $u$ and $d$ quarks in
fig.\ \ref{fig:3dQCDPD}, which I borrowed from
ref.\ \cite{Karsch:2019mbv}.
Since chiral symmetry is exact in the chiral limit, spontaneous
breaking of chiral symmetry must happen through a phase
transition \cite{Halasz:1998qr}. The zero temperature transition
is expected to be of first order and it extends in the $T-\mu_B$ plane
and bends toward the $T$-axis. This is shown by the thick
black line in fig.\ \ref{fig:3dQCDPD}. For vanishing chemical potential,
on the other side, it was argued in a seminal work
\cite{Pisarski:1983ms} that the chiral phase transition for massless
$u$ and $d$ quarks will be of second order belonging to the $O(4)$
universality class (red dot in fig.\ \ref{fig:3dQCDPD}). This also extends for finite $\mu_B$ and bends towards the $\mu_B$-axis
(represented by the thick red line in fig.\ \ref{fig:3dQCDPD}) and
meets the before-mentioned first order line in a tri-critical point (magenta dot in fig.\ \ref{fig:3dQCDPD}). This completes the picture
in the chiral limit and indicates $T_c>T_{\text{tri}}$.
When one leaves the chiral plane, the first
order transition at low $T$ and high $\mu_B$ remains first order
but the transition at vanishing and small but non-vanishing values
of $\mu_B$ becomes a crossover which is represented by the black
dashed line in fig.\ \ref{fig:3dQCDPD}. The first order line in
the $T-\mu_B$ plane for finite quark masses ends in a (bi-)critical
point which shifts to smaller $T$ (so $T_{\text{tri}}>T_{\text{cep}}$) and larger $\mu_B$
\cite{Hatta:2002sj}, which defines the so-called wing line(s),
represented by thick blue line in fig.\ \ref{fig:3dQCDPD}.
The critical end-point at finite masses are connected to the temperature
axis through a crossover, and this crossover temperature at $\mu_B=0$ for
physical pion mass has been recently determined very precisely \cite{HotQCD:2018pds,Borsanyi:2020fev}.
It is not hard to realize that the QCD phase diagram at physical
value of pion mass is nothing but a slice of the more general
phase diagram represented in fig.\ \ref{fig:3dQCDPD}.
These interlinked pieces of information gives rise to a very
important inequality: $T_{\text{cep}}<T_c$, {\it i.e.}\ the chiral
critical temperature puts an upper bound on the position of
the conjectured critical end point in the physical world QCD.

The picture presented in the last paragraph is not undisputed.
In the before-mentioned work by Pisarski and Wilczek \cite{Pisarski:1983ms},
it is also mentioned that if the strength of the anomaly decreases
with increasing temperature and $U_A(1)$ gets effectively restored
at the critical temperature then the chiral transition for
two massless flavors can be of first order. Then the scenario
will be quite different compared to what is shown in fig.\ \ref{fig:3dQCDPD}.
Although later it was proposed \cite{Pelissetto:2013hqa} that
the chiral transition for two massless flavors can still be
continuous even if $U_A(1)$ is restored, but then the
universality class will be different.

\begin{figure}
    \centering
    \includegraphics[scale=0.6]{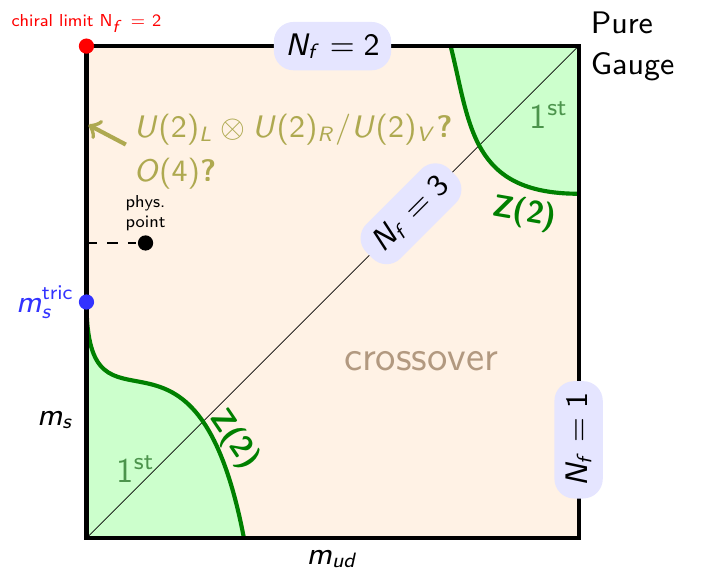} ~~~~
    \includegraphics[scale=0.6]{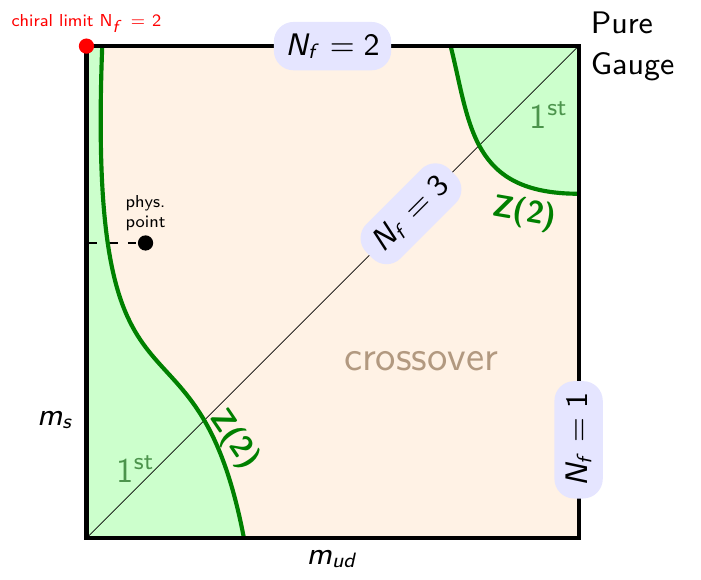}
    \caption{Two most celebrated versions of the Columbia plot.
    Figures are taken from ref.\ \cite{Philipsen:2016hkv}.}
    \label{fig:ColumbiaPlot}
\end{figure}

In this context, two widely celebrated versions of
a flag diagram, known as the Columbia plot \cite{Brown:1990ev}
which presents the order of the chiral transition in the plane
of light (degenerate $u$ and $d$) and strange quark masses,
is shown in fig.\ \ref{fig:ColumbiaPlot}. In the left panel
the second order scenario for $2$- and ($2+1$)-flavor is shown,
whereas the right panel shows the scenario where the chiral
transition is of first order. I will come to more details of
these diagrams later. Various other aspects about the
Columbia plot can be found in \cite{Gupta:2008ac}.

I hope by now I convinced the reader that why/how the calculation
of the chiral critical temperature and determination of the order
of the chiral phase transition is of fundamental necessity.
In this talk I shall try to map the progress made by the community
to shed light on various aspects of QCD towards the chiral limit.
My apologies if I miss any reference; it goes without saying
that would be completely unintentional.

\section{Results: a few mentions}

In this part I shall try to mention various works that have tried
and are trying to answer various fundamental questions about the
chiral phase transition. First I mention the calculations regarding
the chiral transition temperature or to determine the order of the
chiral transition for two degenerate light flavors ($N_f=2$), and then I mention
the same for three ($N_f=3$) and larger than three ($N_f>3$) numbers
of degenerate flavors. Finally I come back to $N_f=2(+1)$ and mention
very recent developments regarding the energy-like observables.

\subsection{\boldmath$N_f=2(+1)$: the chiral transition temperature}

Let me start with a well known scaling formula for the
pseudo-critical temperatures, defined by the peak position
of order parameter susceptibility for various quark masses:

\begin{equation}
    T_{\text{p}}(H) = T_c \left( 1 + \frac{z_p}{z_0} H^{1/\beta\delta} \right) + \text{sub-leading} \; ,
    \label{eq:peakscaling}
\end{equation}
where $T_p(H)$ is the peak position for the symmetry breaking field
$H$, which in QCD is proportional to the light quark masses. The chiral
transition temperature $T_c$
and the scale of the scaling variable $z_0$ are non-universal
parameters, and $z_p$ specifies the peak position of the scaling
function of the order parameter susceptibility. For the relevant
universality classes regarding the chiral transition for two massless
flavors, the value of $z_p$ is order unity \cite{HotQCD:2019xnw} which
gives rise to a significant drop in a non-linear fashion of the pseudo-critical
temperature towards the chiral limit, w.r.t.\ {\it e.g.}\ physical
values of the light quark masses. Authors of \cite{HotQCD:2019xnw}
came up with two novel estimators of pseudo-critical temperatures
by the following conditions;

\begin{equation}
    \frac{H\chi_M(T_{\delta}),H}{M(T_{\delta},H)}=\frac{1}{\delta}
    ~~~~ \text{and} ~~~~ \chi_M(T_{60},H)=0.6~\chi_M(T_p,H) \; ,
    \label{eq:improvedestimators}
\end{equation}
where $M$ and $\chi_M$ are the order parameter (proportional to the
chiral condensate) and its susceptibility (proportional to the chiral
susceptibility) for the chiral phase transition. The values of the scaling
variable corresponding to the conditions in
eq.\ \ref{eq:improvedestimators}, named as $z_{\delta}$ and $z_{60}$,
are two orders of magnitude smaller compared to $z_p$
\cite{HotQCD:2019xnw}, which results in way less variation of the
pseudo-critical temperatures towards the chiral limit.

\begin{figure}[!h]
    \centering
    \includegraphics[scale=0.25]{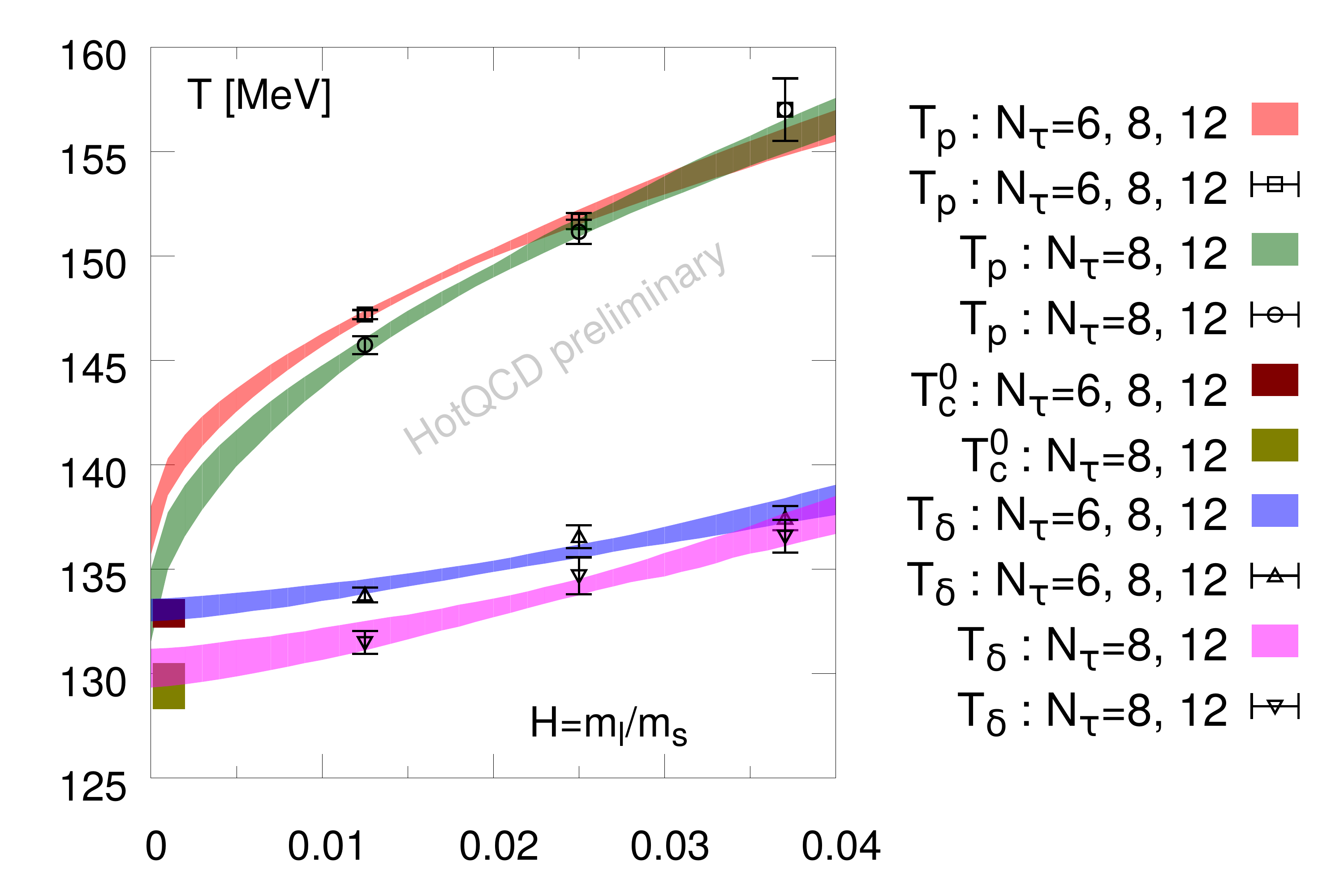} ~~~~
    \includegraphics[scale=0.40]{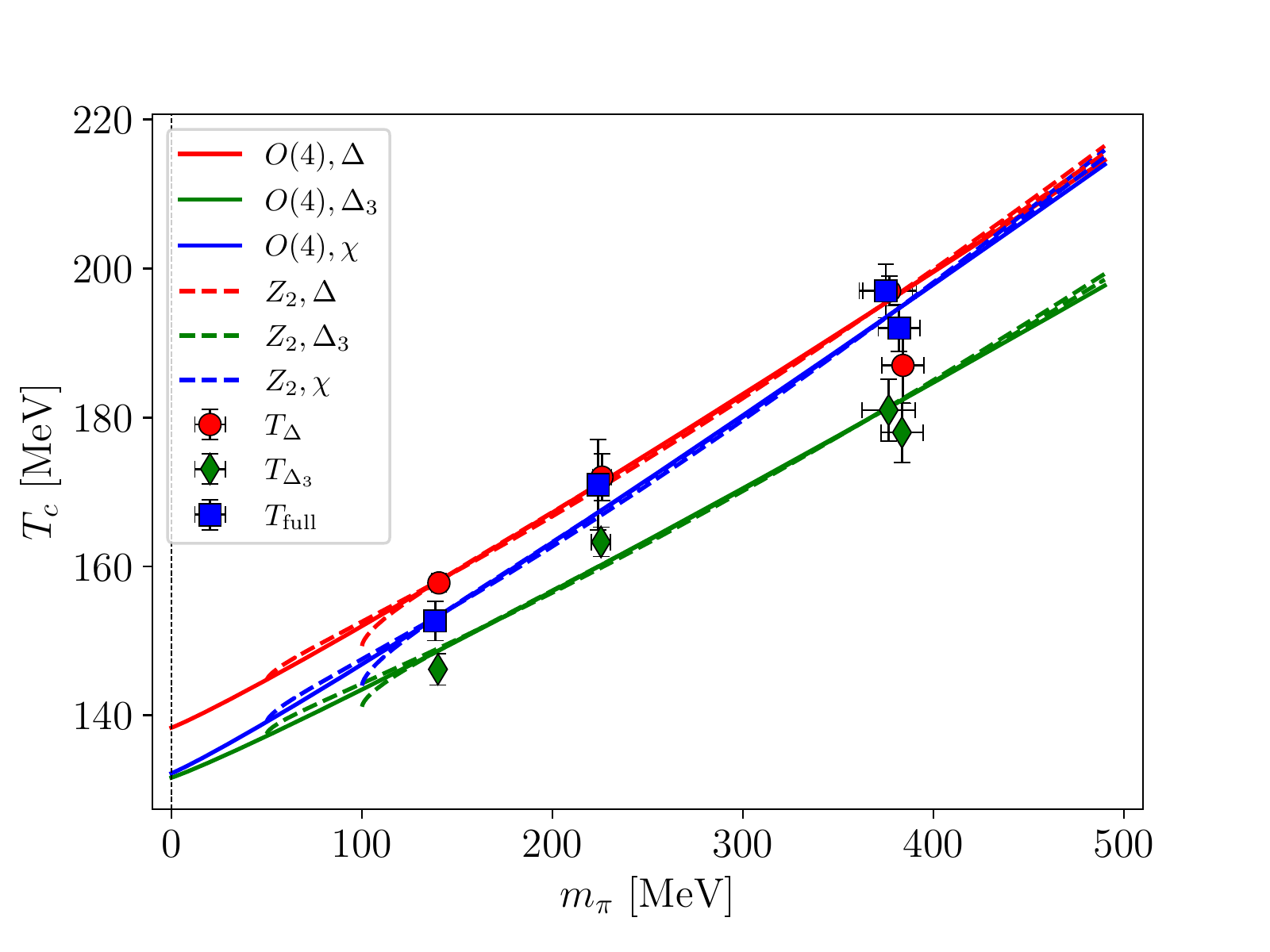}
    \caption{Chiral extrapolation of various pseudo-critical
    temperatures. Left: calculation with HISQ and the plot is taken
    from \cite{Kaczmarek:2020err}. Right: calculation with twisted mass
    Wilson fermion and the plot is taken from \cite{Kotov:2021rah}.}
    \label{fig:Tpcextrapolation}
\end{figure}

In fig.\ \ref{fig:Tpcextrapolation} the chiral extrapolation of various
pseudo-critical temperatures from two different groups is shown. The left panel
shows the {\it continuum extrapolated} results of $T_{\delta}$ and
$T_{60}$, and they are compared with the conventional estimator of $T_p$
({\it also continuum extrapolated}), calculated with highly improved staggered quarks (HISQ). The stability of the new estimators
is vivid in the plot. In ref.\ \cite{HotQCD:2019xnw} it is found that
the behavior of various pseudo-critical estimators is consistent
with $O(4)$ scaling ans\"{a}tze. Finally the chiral critical temperature,
{\it in the continuum}, is quoted as $T_c=132^{+3}_{-6}$ MeV \cite{HotQCD:2019xnw}. In the right
panel the chiral extrapolation of various pseudo-critical estimators,
calculated with twisted mass Wilson fermions \cite{Kotov:2021rah}, are shown. In contrary to
ref.\ \cite{HotQCD:2019xnw}, this calculation is performed within
a fixed scale approach. Still the quoted chiral critical temperature,
$T_c=134^{+6}_{-4}$ MeV, agrees well with the other calculation.
Consistency with $O(4)$ scaling is found. Moreover, the chiral
extrapolations of pseudo-critical estimators are found not quite sensitive
to the difference on the universality classes.

\subsection{\boldmath$N_f=2(+1)$: order of the chiral transition}
Coming to the nature of the chiral transition for $N_f=2$, there exist
two major ways to look into it: either from the scaling perspective or
through direct determination of the fate of $U_A(1)$. In the following
I will take the tour in both ways.

\subsubsection{\boldmath $N_f=2(+1)$: order of the chiral transition from scaling}

\begin{figure}[!h]
    \centering
    \includegraphics[scale=0.55]{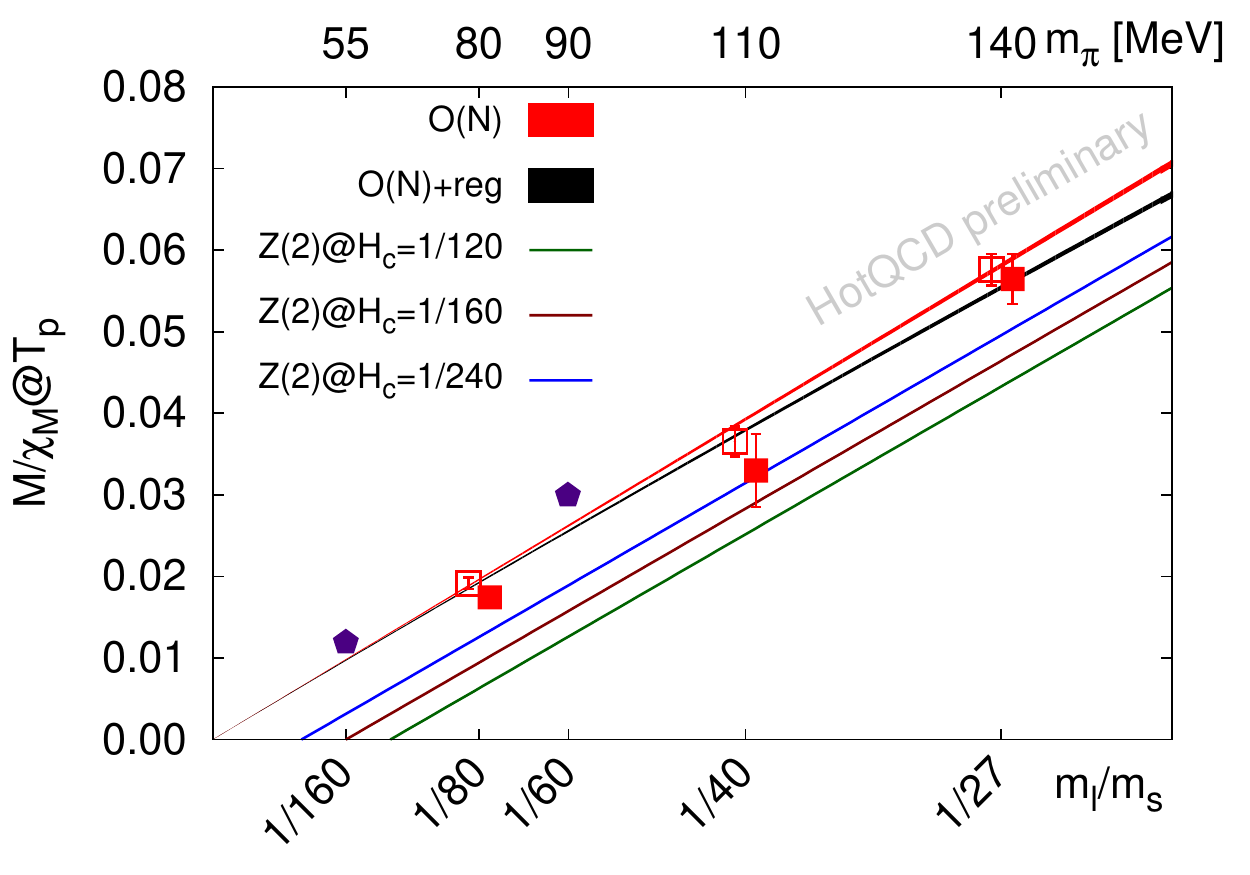} ~~~~
    \includegraphics[scale=0.77]{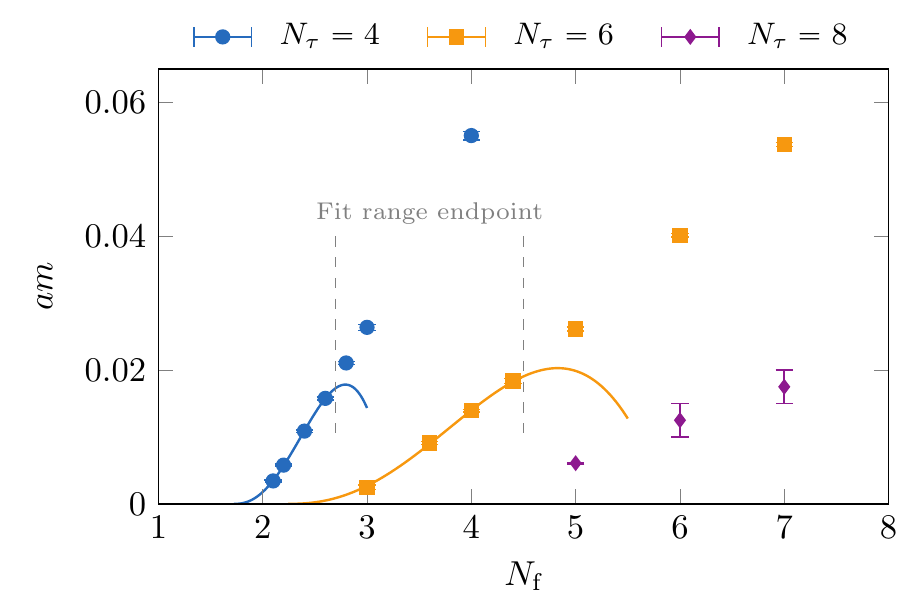}
    \caption{Left: Continuum extrapolated ratio of the chiral order
    parameter and its susceptibility evaluated at the peak of the chiral
    susceptibility $T_p$ plotted as function of the scaled light quark
    masses. Plot is taken from \cite{Kaczmarek:2020err}.
    Right: Bare critical quark mass at which a second order phase
    transition belonging to $Z_2$ universality class is found, shown
    as a function of number of flavors $N_f$ for various $N_{\tau}$.
    Plot is taken from \cite{Cuteri:2021ikv}.}
    \label{fig:chiralPTorder}
\end{figure}

Close to the critical point the ratio of the order parameter and its
susceptibility is given by \cite{Kaczmarek:2020err}
\begin{equation}
    \left.\frac{M}{\chi_M}\right|_{T_X,H}=\left.\frac{f_G(z)}{f_{\chi}(z)}\right|_{z_X} H ~+~ \text{sub-leading}.
    \label{eq:ratiounivclass}
\end{equation}
where $X=\text{p},~\delta$ or 60.
It can be clearly seen that in absence of the sub-leading corrections
the RHS is determined solely through the universal contribution.
This fact has been exploited in the left panel of
fig.\ \ref{fig:chiralPTorder}. Parameter free comparison
of the data to the scaling expectation clearly depicts that the
data prefers an $O(N)$ scenario ($O(4)$ in continuum and $O(2)$ for
staggered calculation at a finite temporal extent, $N_{\tau}$) over
a first order transition in the chiral limit. Little deviations of the
data w.r.t.\ the scaling expectation towards larger quark masses can be
accommodated using a regular term, which is proportional to quark mass
in the order parameter. Similar results are found for $z_{60}$. More 
details about this analysis can be found in \cite{Kaczmarek:2020err}.

There is another attempt to determine the order of the chiral transition
using tri-critical scaling. In ref.\ \cite{Cuteri:2021ikv} the number
of degenerate quark flavors was treated as a real continuous parameter
and found the phase boundary in an extended parameter space. In the right
panel of fig.\ \ref{fig:chiralPTorder} only the projection in the plane
of quark mass and number of flavors is shown. Each data point implicitly
specifies a critical coupling. The chiral transition is first order
below this set of points for a given $N_{\tau}$  and above
it is crossover. Each of these critical points corresponds to second order
phase transitions belonging to the $Z_2$ universality class. These are usually
calculated on the basis of the finite size scaling arguments of the kurtosis of
the chiral condensate. These critical points, creating a line, are eventually
expected to go to a tri-critical point at some critical value of $N_f$,
as can be realized from the left panel of fig.\ \ref{fig:ColumbiaPlot}
comparing the situation for $N_f=3$ and 2. In the right panel of
fig.\ \ref{fig:chiralPTorder}, one can clearly see the
sizeable shift of the critical value of $N_f$ w.r.t.\ $N_{\tau}$, and
it is quite clear that in the continuum the critical $N_f$ will be
larger than 2, implying that the chiral transition for $N_f=2$ will be
of second order.

\subsubsection{\boldmath$N_f=2(+1)$: effective restoration of $U_A(1)$}

\begin{figure}
\centering
\includegraphics[scale=0.5]{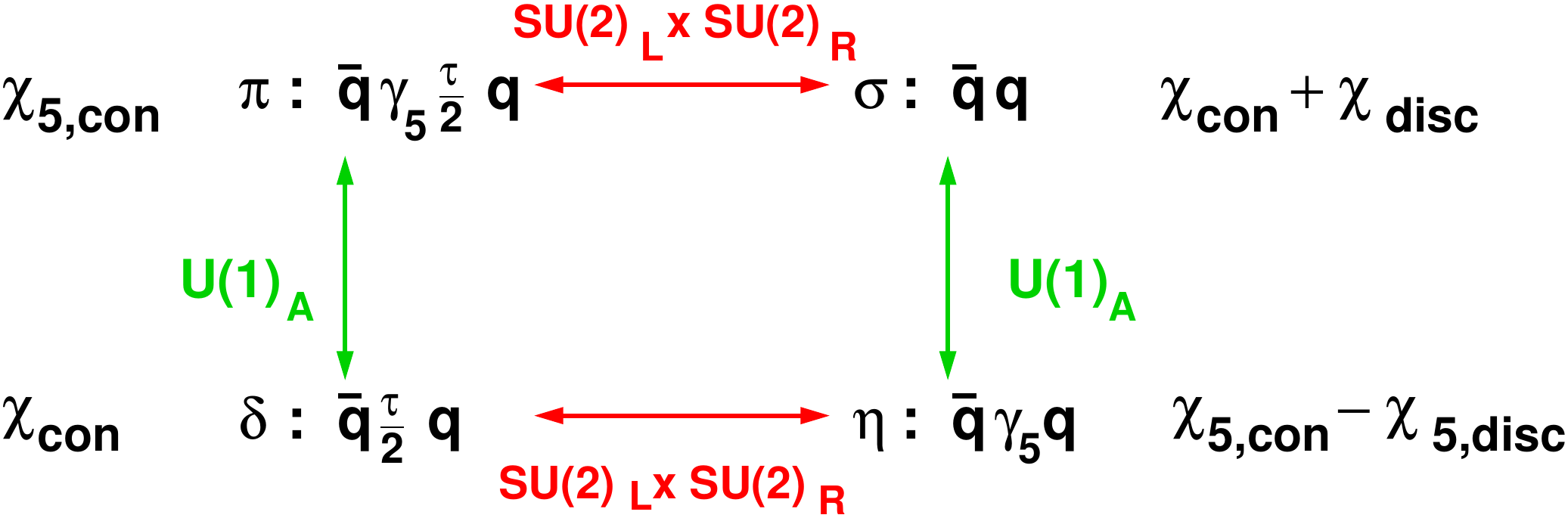}
\caption{Various mesonic channels of QCD and their relation through
either the flavor non-singlet chiral transformation or by
the flavor singlet axial transformation.
Figure is taken from ref.\ \cite{HotQCD:2012vvd}.}
\label{fig:symmetry}
\end{figure}

The $U_A(1)$ symmetry, which is broken in nature upon quantization,
is expected to get effectively restored at high temperature because
of the deceasing number of non-perturbative topological configurations
\cite{Pisarski:1983ms}. So investigation regarding the effective
restoration of $U_A(1)$ is extremely important to establish the
nature of the chiral transition for 2-flavor and (2+1)-flavor QCD.
Before I go to mentioning different studies in this direction, let
me start with a brief introduction how effective
restoration of $U_A(1)$ is usually studied in the literature. In this context
I borrowed fig.\ \ref{fig:symmetry} from ref.\ \cite{HotQCD:2012vvd}.
In this schematic diagram various mesonic channels are connected
through either the flavor non-singlet chiral transformation or by
the flavor singlet axial transformation, $U_A(1)$. The well known
expected degeneracy between iso-triplet pseudo-scalar meson ($\pi$) and
iso-scalar scalar meson ($\sigma$ or $f_0$) can be seen through
the horizontal connection between the top two entries of fig.\ \ref{fig:symmetry}.
Similarly the iso-triplet pseudo-scalar meson ($\pi$) is related to
the iso-triplet scalar meson ($\delta$ or $a_0$) by the flavor singlet
axial transformation, $U_A(1)$, which is represented by the vertical
connection between the two left side entries of fig.\ \ref{fig:symmetry}.
To investigate the effective restoration of $U_A(1)$, usually
the degeneracy between the masses or susceptibilities is checked,
both of which is actually based upon the degeneracy between the correlation
functions in those channels. Specifically most of the literature
calculates differences like $m_{a_0/\delta}-m_{\pi}$ or
$\chi_{\pi}-\chi_{a_0/\delta}$ as a measure of the $U_A(1)$
breaking.

It is interesting to note that in the {\it chirally symmetric background}
one can use the degeneracy between the diagonal entries of
fig.\ \ref{fig:symmetry} to check the effective
restoration of $U_A(1)$. When one chooses to connect the diagonal
containing $\delta$ and $\sigma$, it can be immediately realized that
the corresponding susceptibilities are given by the connected
part and the total chiral susceptibility. As a result of this
identification one can write \cite{Kaczmarek:2020sif}
\begin{eqnarray}
    \chi_{\pi}-\chi_{a_0/\delta}&=&\left(\chi_{\pi}-\chi_{\sigma}\right) +
    \chi_{\text{disc}} \nonumber \\
    &=&\chi_{\text{disc}}, ~~ \text{in~the~chirally~symmetric~phase,}
\end{eqnarray}
which implies the condition for the effective restoration of $U_A(1)$
boils down to the condition whether in the chirally symmetric
background $\chi_{\text{disc}}$ vanishes or not. 

A plethora of literature can be found about the effective restoration of
$U_A(1)$ and is not possible to cover all of those here. In the following
I rather list some of the recent efforts which are almost equally divided
into the groups which either favors or disfavors the effective restoration
of $U_A(1)$. I show one key plot from each study and mention few
features of the calculations belonging to either world.

\begin{figure}
    \centering
    \includegraphics[scale=0.35,trim={1cm 0 0 0}]{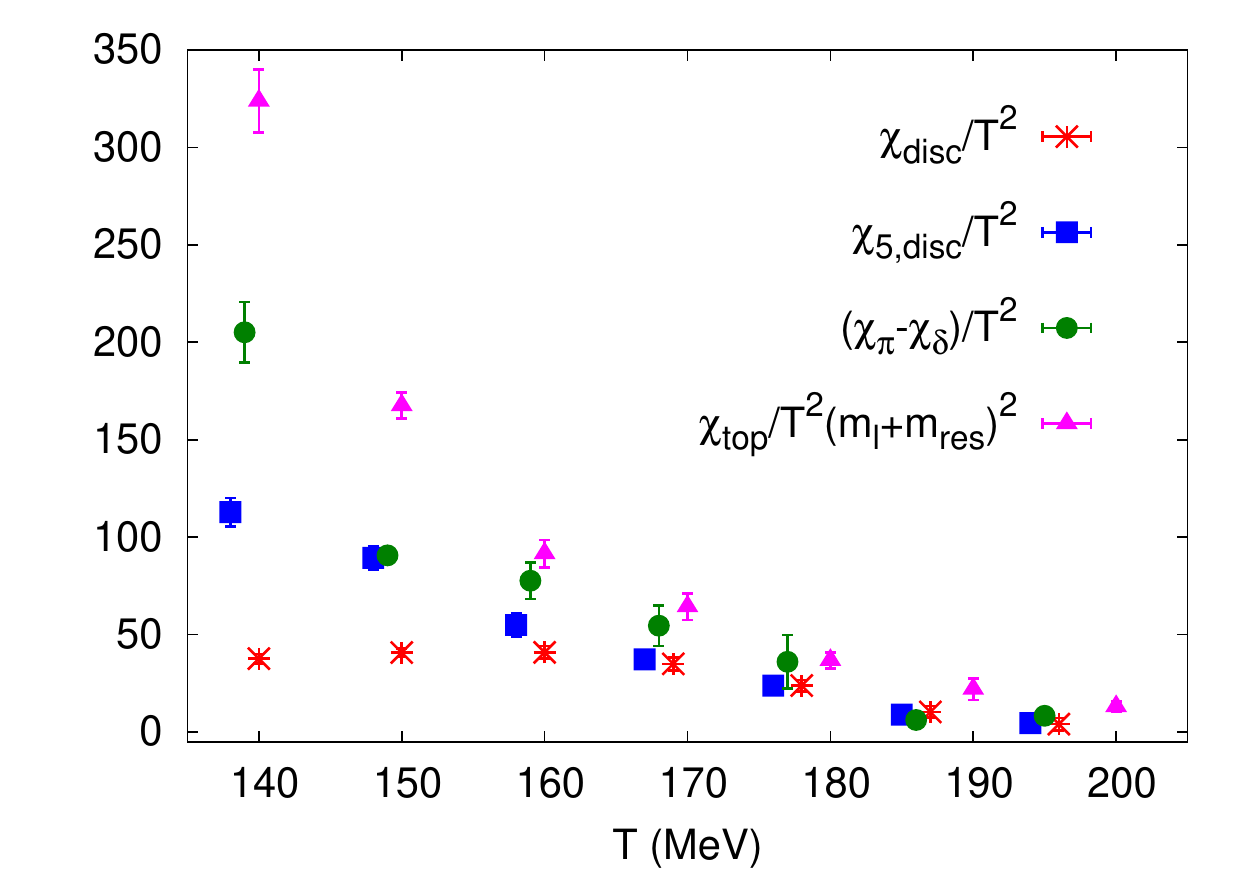} ~~~~
    \includegraphics[scale=0.35,trim={1cm 0 0 0}]{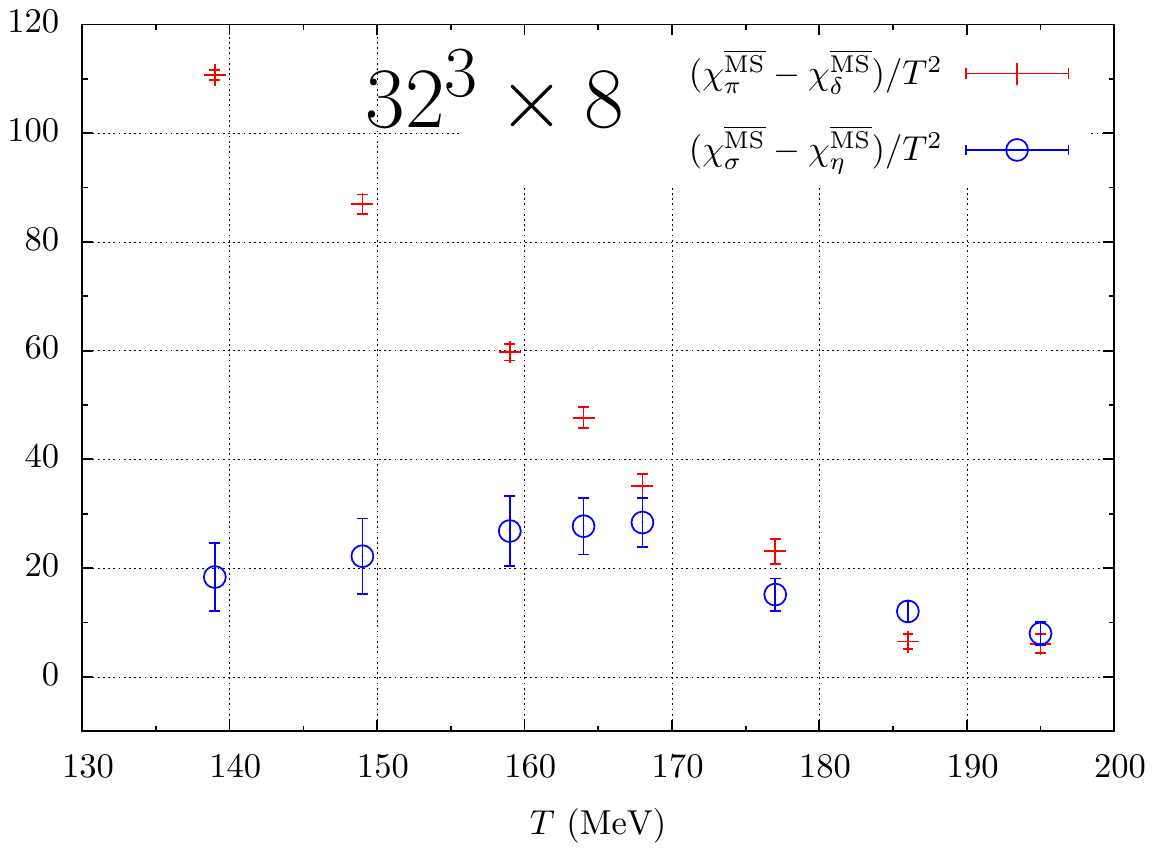} ~~~~
    \includegraphics[scale=0.29,trim={0 0 0 0}]{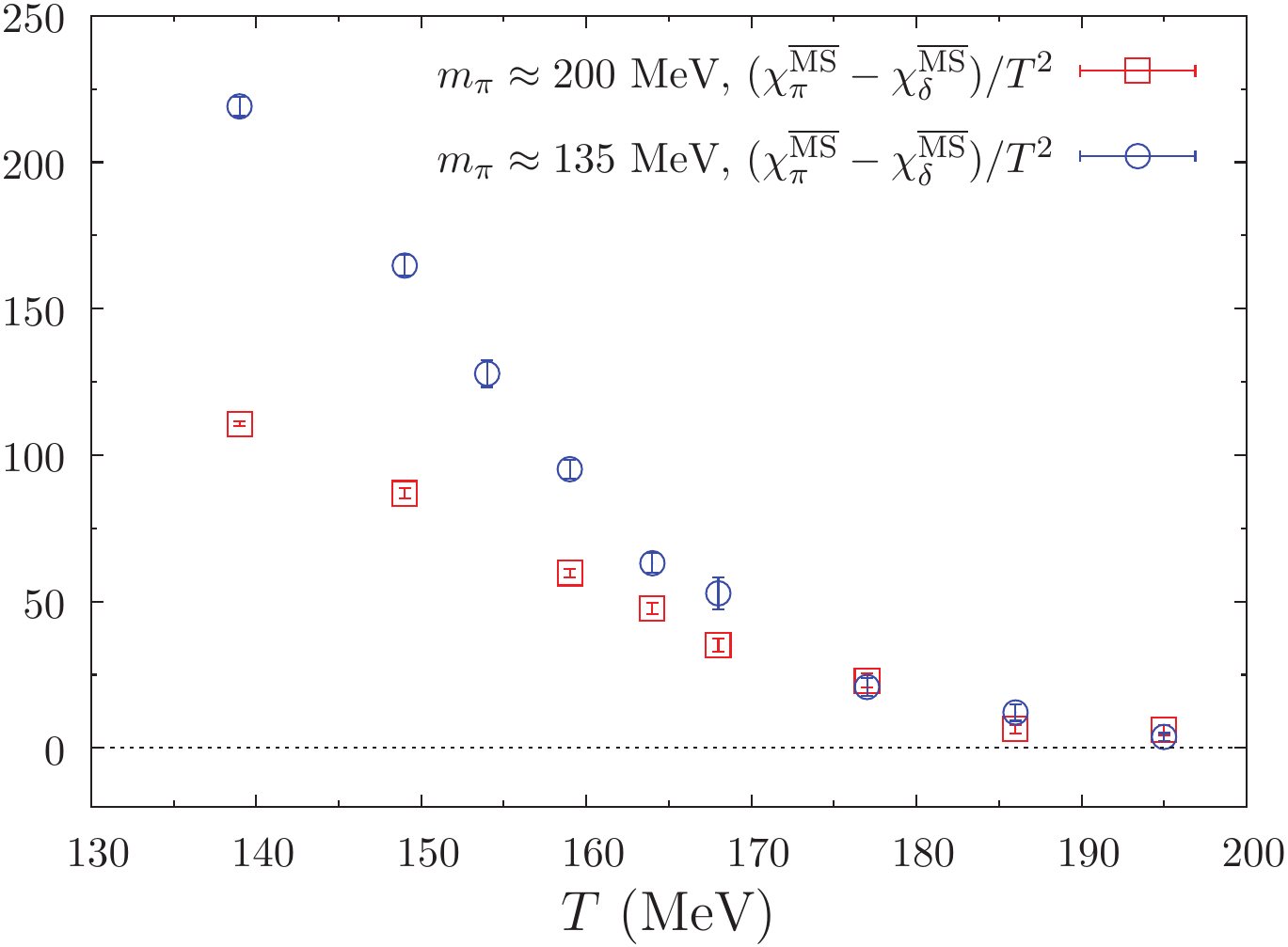}
    \caption{Temperature variation of symmetry breaking measures from calculation with
             (2+1)-flavor domain-wall fermions (DWF) for $N_{\tau}=8$. Plot taken from -
             Left:  ref.\ \cite{HotQCD:2012vvd},
             Middle: ref.\ \cite{Buchoff:2013nra},
             Right: ref.\ \cite{Bhattacharya:2014ara}.}
    \label{fig:UA1_DWF_HotQCD}
\end{figure}

In the left panel of Fig.\ \ref{fig:UA1_DWF_HotQCD} the temperature variation
of symmetry breaking measures is shown from a calculation with
(2+1)-flavor domain-wall fermions (DWF) for $N_{\tau}=8$
\cite{HotQCD:2012vvd} with a $\sim$ 45\% heavier pion compared to the 
physical one. $\chi_{\pi}-\chi_{\delta}$
remains non-zero for the entire temperature interval of the study
suggesting that $U_A(1)$ is effectively broken even after restoration
of chiral symmetry. Moreover degeneracy of $\chi_{\pi}-\chi_{\delta}$
with the disconnected scalar/chiral susceptibility $\chi_{\text{disc}}$
and disconnected pseudo-scalar susceptibility $\chi_{5,\text{disc}}$ 
at high temperatures ensures that the non-vanishing
$\chi_{\pi}-\chi_{\delta}$ originates indeed from $U_A(1)$ breaking
and not from explicit breaking due to quark masses.

\begin{wrapfigure}{r}{0.4\textwidth}
    \centering
    \includegraphics[scale=0.2]{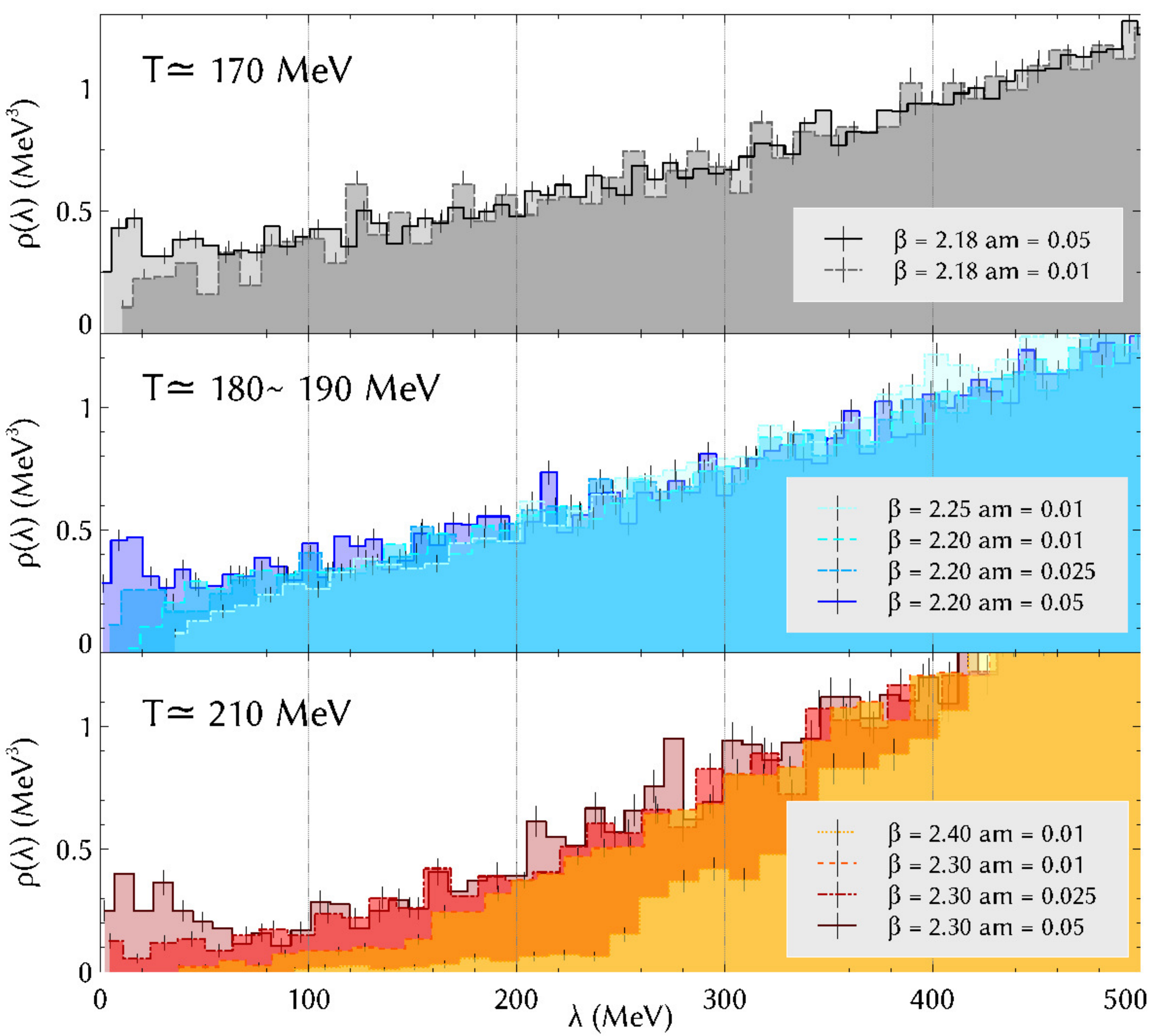}
    \caption{Eigenvalue spectrum of the massless 2-flavor overlap Dirac operator for
             $N_{\tau}=8$ \cite{Cossu:2013uua}. Lighter shed represents smaller masses.}
    \label{fig:UA1_OV_2013}
\end{wrapfigure}

A follow-up study \cite{Buchoff:2013nra} with the same set up as
above but with larger volumes confirms that the finite volume effects
are under control. As can be seen from the middle panel of fig.\ \ref{fig:UA1_DWF_HotQCD}
the $U_A(1)$ breaking measures have a clearly non-vanishing value,
and the near equality between the different measures ensures
that the explicit chiral symmetry breaking is tiny around the
highest temperatures of this study. Further analyses of
the volume dependence and the chirality of the near-zero modes
agrees with the expectations from the dilute instanton gas approximation
picture, again confirming breaking of $U_A(1)$ above the pseudo-critical
temperature.

In the right panel of fig.\ \ref{fig:UA1_DWF_HotQCD} the
temperature variation of $\chi_{\pi}-\chi_{\delta}$ from calculation
with (2+1)-flavor domain-wall fermions (DWF) for $N_{\tau}=8$
\cite{Bhattacharya:2014ara} with a physical pion is shown. Again the $U_A(1)$
breaking measure remains non-zero over the entire temperature range of
the study. At low temperature the expected increase of
$\chi_{\pi}-\chi_{\delta}$ with decreasing quark mass can be seen.
The apparent mass independence of $\chi_{\pi}-\chi_{\delta}$ at high
temperatures confirms that the non-vanishing value comes due to
the breaking of $U_A(1)$ and not due to small explicit breaking through
non-vanishing quark masses.

In fig.\ \ref{fig:UA1_OV_2013} the eigenvalue spectrum of
the massless 2-flavor overlap Dirac operator for $N_{\tau}=8$
\cite{Cossu:2013uua} is shown. Lighter sheds represent smaller masses.
The upper panel corresponds to $T<T_{\text{pc}}$,
where reduction of the mass reduces the density of near zero modes $\rho(0)$,
which is compatible with finite volume effects. In the infinite volume
limit $\rho(0)$ is expected to be finite. The middle panel corresponds to
$T\sim T_{pc}$, and here decreasing the mass resulted in suppression
of eigenvalues up to $\sim 40$ MeV, which is consistent with a vanishing
$\rho(0)$ in the chiral limit. This suppression is even more
prominent for $T>T_{\text{pc}}$ and indicates towards a gap in the
spectrum in the chiral limit. Such a gapped spectrum eventually leads
to the degeneracy between correlators in $\pi$ and $\delta$ channels,
implying effective restoration of $U_A(1)$ in the chiral limit.

\begin{figure}
    \centering
    \includegraphics[scale=0.40]{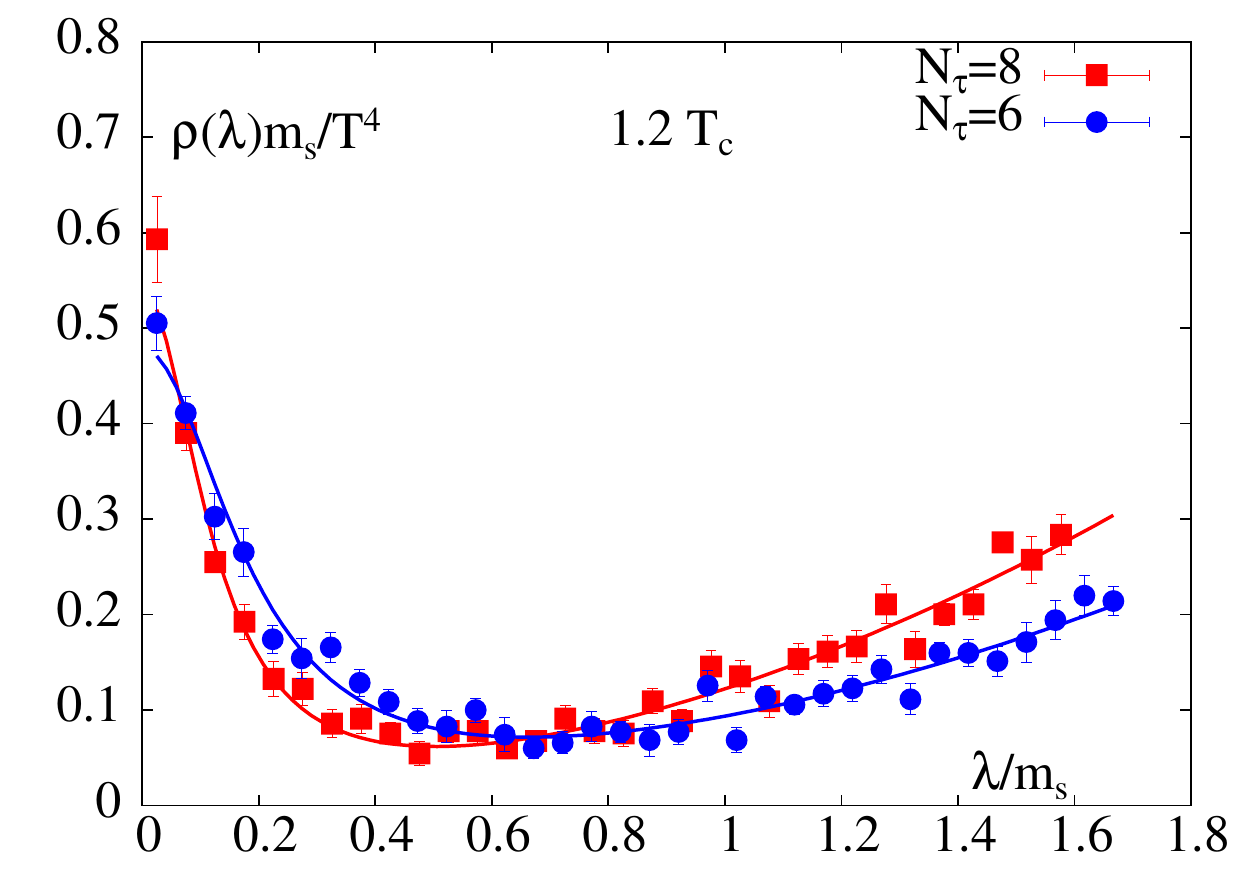} ~~~~
    \includegraphics[scale=0.60]{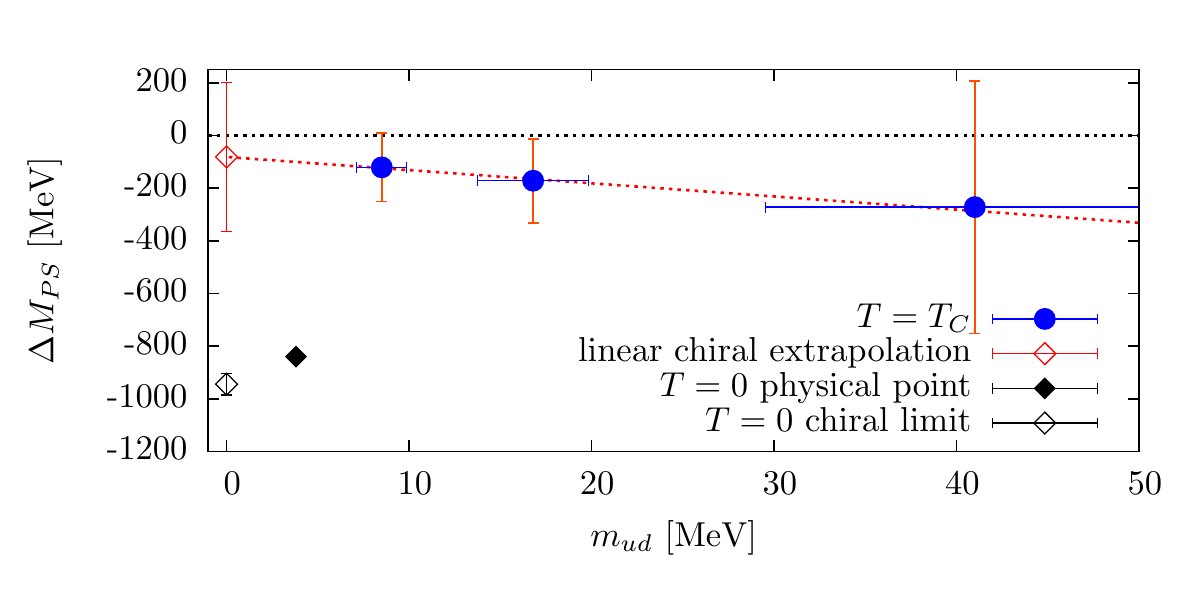}
    \caption{Left: Renormalized eigenvalue spectra calculated with overlap fermions
             on (2+1)-flavor HISQ ensembles \cite{Dick:2015twa} for $N_{\tau}=6,8$.
             Right: Chiral extrapolation of the difference between the pseudo-scalar and
             scalar masses, calculated using non-perturbatively ${\cal O}(a)$-improved
             2-flavor Wilson fermions \cite{Brandt:2016daq} with $N_{\tau}=16$.}
    \label{fig:UA1_leftHISQ_rightWilson}
\end{figure}

\begin{wrapfigure}{l}{0.4\textwidth}
    \includegraphics[scale=0.55]{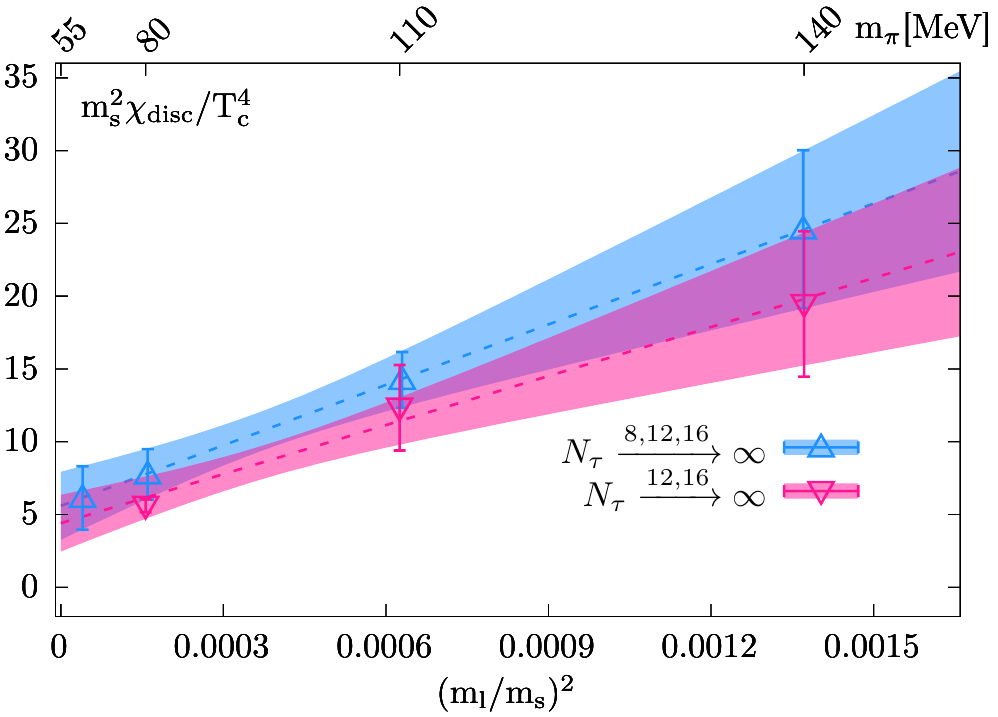}
    \caption{Chiral extrapolation of $\chi_{\text{disc}}$ calculated using (2+1)-flavor HISQ
     action with $N_{\tau}=8,~12$ and 16, for $T=1.6~T_c$ \cite{Ding:2020xlj}.}
    \label{fig:UA1_HISQ_cont}
\end{wrapfigure}

In the left panel of fig.\ \ref{fig:UA1_leftHISQ_rightWilson} renormalized eigenvalue
spectra calculated with overlap fermions on (2+1)-flavor HISQ
ensembles \cite{Dick:2015twa} for $N_{\tau}=6,8$ are shown.
Accumulation of near-zero modes being independent of the cutoff
and being stable under smearing of the gauge fields
ensures that the accumulation is not due to lattice artifacts.
Since these near-zero modes will result into a non-vanishing
$\chi_{\pi}-\chi_{\delta}$, apparently $U_A(1)$ seems to be broken
even above the pseudo-critical temperature. Detailed study of the
spatial structure and localization properties of these near-zero 
modes reveals agreement with the expectation from weakly interacting
instantons and anti-instantons.

In the right panel of fig.\ \ref{fig:UA1_leftHISQ_rightWilson} a chiral extrapolation
of the difference between the pseudo-scalar and scalar masses at
the pseudo-critical temperature of the respective masses is shown,
calculated using non-perturbatively ${\cal O}(a)$-improved 2-flavor
Wilson fermions \cite{Brandt:2016daq} with $N_{\tau}=16$.
Extrapolations with linear and square-root ans\"{a}tze w.r.t.\ quark
masses lead to a vanishing value (within uncertainties) of the mass
difference shown in the plot, implying an effective restoration of
$U_A(1)$ in the chiral limit.

In fig.\ \ref{fig:UA1_HISQ_cont} chiral extrapolation of $\chi_{\text{disc}}$
calculated using the (2+1)-flavor HISQ action with $N_{\tau}=8,~12$ and 16 is
shown for $T=1.6~T_c$, where $T_c$ is the chiral critical temperature in
the continuum \cite{Ding:2020xlj}. The blue band
represents a joint chiral and continuum extrapolation using an ans\"{a}tz
that is quadratic in quark mass, and the coefficients have a quartic correction
in lattice spacing. The magenta band, on the other hand, shows the result from the
\textquoteleft proper\textquoteright\ order of limits for staggered calculations where
the continuum limit is taken first with a quadratic correction in lattice spacing
using the two largest $N_{\tau}$ and then mass extrapolation using a quadratic
ans\"{a}tz. Both extrapolations lead to a non-vanishing value of
$\chi_{\text{disc}}$ with more than 95\% confidence, implying that $U_A(1)$
is broken in continuum even at $1.6~T_c$. Similar results are obtained
from the analysis of $\chi_{\pi}-\chi_{\delta}$. Analyses of the correlation
function of the eigenvalue density reveals that the microscopic origin
of the axial anomaly at high temperature can be described within
the weakly interacting (quasi)instanton gas picture.
A similar analysis at somewhat lower temperature, closer to $T_{\text{pc}}$,
on the other hand shows that the eigenvalue spectrum is quite different
from that at higher temperatures and is not consistent with the dilute
instanton gas picture \cite{Ding:2021gdy}.

\begin{wrapfigure}{r}{0.5\textwidth}
\centering
    \includegraphics[scale=0.7,trim={1cm 0 0 0}]{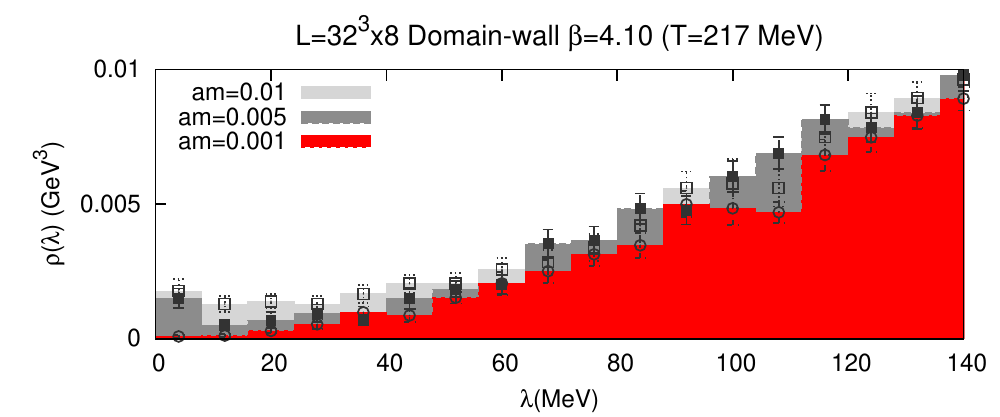}
    \includegraphics[scale=0.7,trim={1cm 0 0 0}]{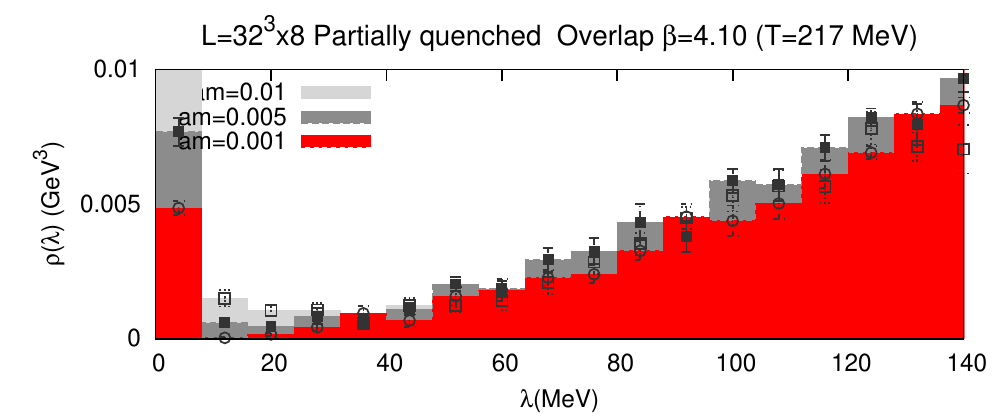}
    \includegraphics[scale=0.7,trim={1cm 0 0 0}]{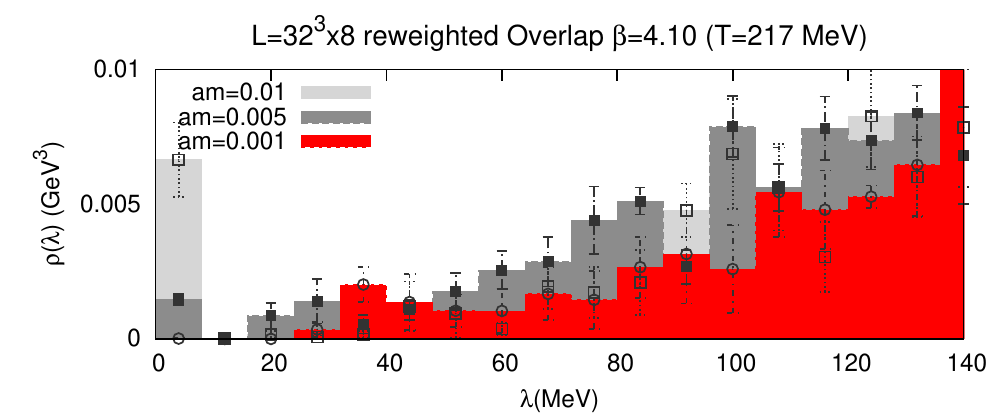}
    \caption{Eigenvalue histograms of domain-wall (top), partially quenched
     overlap (middle) and reweighted overlap (bottom) Dirac operators on
     ensembles generated with 2-flavor domain-wall sea quarks \cite{Tomiya:2016jwr}.}
     \label{fig:UA1_OV_2016}
\end{wrapfigure}

In fig.\ \ref{fig:UA1_OV_2016} eigenvalue histograms of domain-wall
(top panel), partially quenched overlap (middle panel) and reweighted overlap
(bottom panel) Dirac operators on ensembles generated with 2-flavor
domain-wall sea quarks \cite{Tomiya:2016jwr}
are shown. The apparent suppression of
near-zero modes for domain-wall and reweighted overlap spectra towards
the chiral limit seems to be stable under changing
lattice spacing and spatial volume implying that the observed effect
is not a lattice artifact. Eigenvalues in the lowest bin monotonically
decrease w.r.t.\ decreasing quark mass and become consistent with zero
in the chiral limit. A highly contrasting behavior can be observed
for the partially quenched overlap spectrum - a sharp peak is present in
the lowest bin and persists against the decrease of quark mass. Since
this peak does not appear in the domain-wall and reweighted overlap 
spectra, this was considered as an artifact of partial quenching.
After removing this artifact $\chi_{\pi}-\chi_{\delta}$ is found to
be consistent with zero in the chiral limit, implying effective
restoration of $U_A(1)$ in the chiral limit.

\begin{figure}[!t]
    \centering
    \includegraphics[scale=0.50,trim={0 0 0 0}]{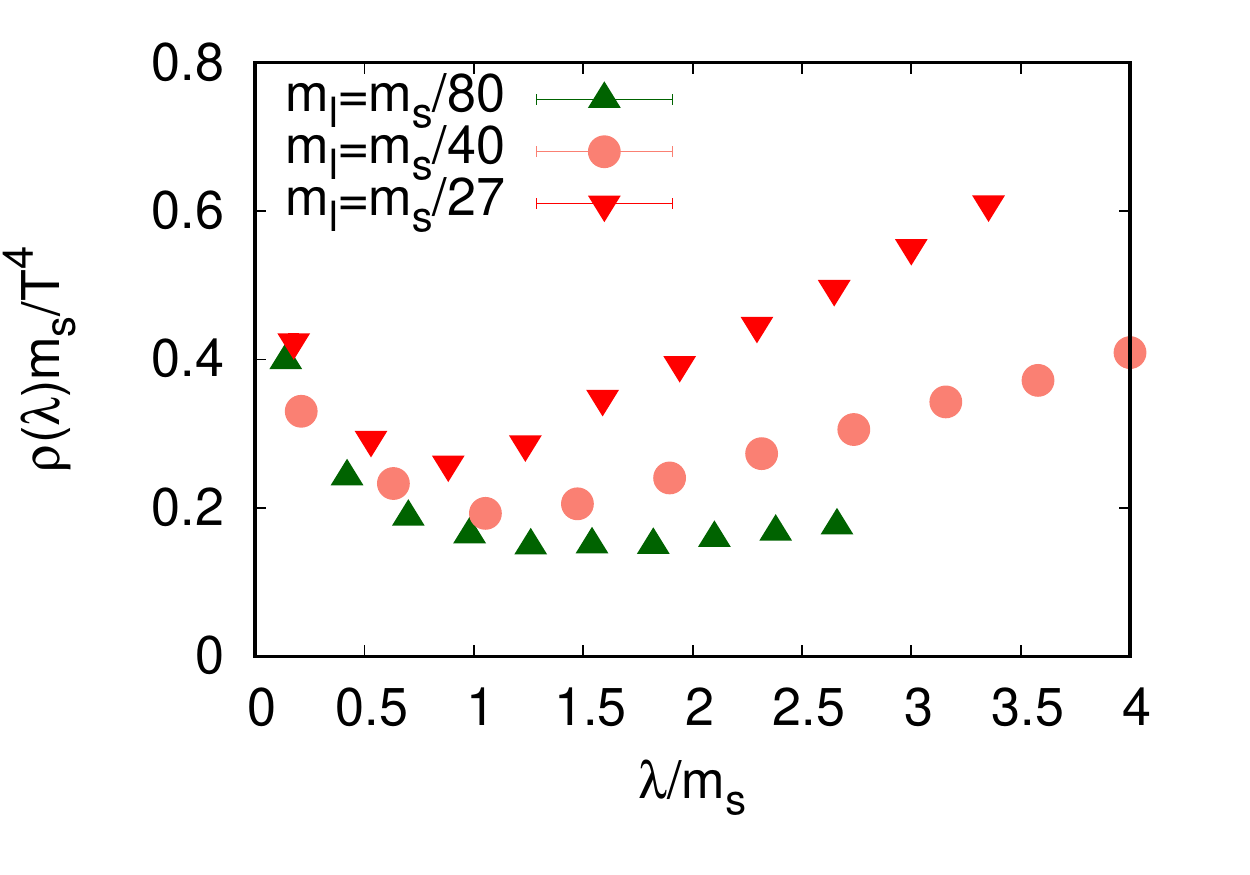} ~~~~
    \includegraphics[scale=0.50,trim={0 0 0 0}]{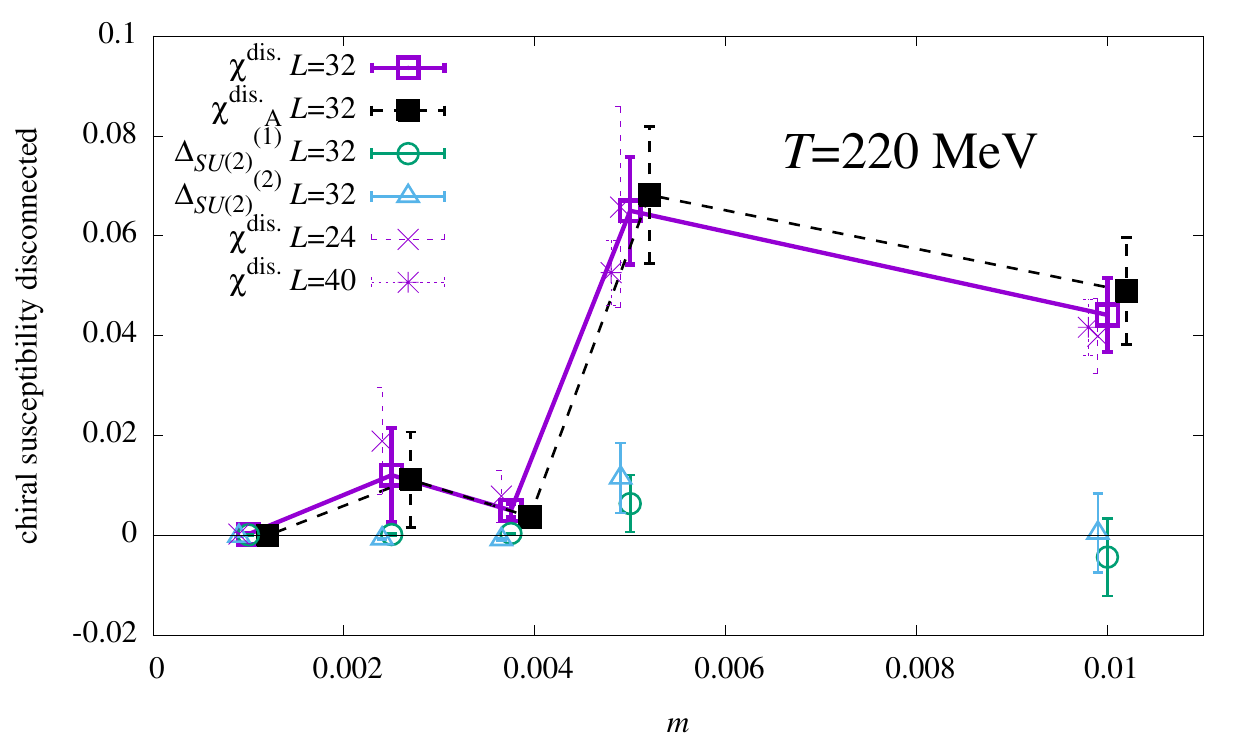}
    \caption{Left: Renormalized eigenvalue density of the appropriately mass-tuned valence
    overlap operator on (2+1)-flavor HISQ ensembles
    for three different light quark masses \cite{Kaczmarek:2021ser}.
    Right: Disconnected part of the chiral susceptibility from a calculation 
    from reweighted overlap Dirac spectrum on M\"{o}bius
    domain-wall sea quarks, within a fixed scale approach \cite{Aoki:2021qws}.}
    \label{fig:UA1_leftHISQ_rightOV}
\end{figure}

In the left panel of fig.\ \ref{fig:UA1_leftHISQ_rightOV}
renormalized eigenvalue densities of appropriately mass-tuned valence
overlap operator on (2+1)-flavor HISQ ensembles
for three different light quark masses \cite{Kaczmarek:2021ser} are shown.
The occurrence of a near-zero peak seems to be stable under a change of quark mass.
The calculation of the renormalized $U_A(1)$ breaking measure using these
eigenvalues confirms that in the chiral limit $U_A(1)$ remains broken
just above the transition temperature.

In the right panel of fig.\ \ref{fig:UA1_leftHISQ_rightOV}
the disconnected part of the chiral susceptibility, obtained from a calculation 
of the reweighted overlap Dirac spectrum on M\"{o}bius domain-wall sea quarks
within a fixed scale approach \cite{Aoki:2021qws}, is shown.
$\chi_{\text{disc}}$ is consistent with zero for the lowest quark mass
and no significant volume dependence is seen, leading to the conclusion
that $U_A(1)$ is restored in the chiral limit.

One important thing to note from all the studies mentioned above is that 
{\it most} of the those probe the temperature range above the pseudo-critical
temperature $T_{\text{pc}}$ and consequently well above the chiral critical
temperature $T_c$. One has to keep in mind that at high temperatures the
$U_A(1)$ breaking anyway becomes small being consistent with the dilute
instanton gas approximation. The effective restoration or breaking of 
$U_A(1)$ thus needs to be checked close to the chiral critical
temperature while going towards the chiral limit. This is difficult!

\begin{figure}[!t]
    \centering
    \includegraphics[scale=0.70]{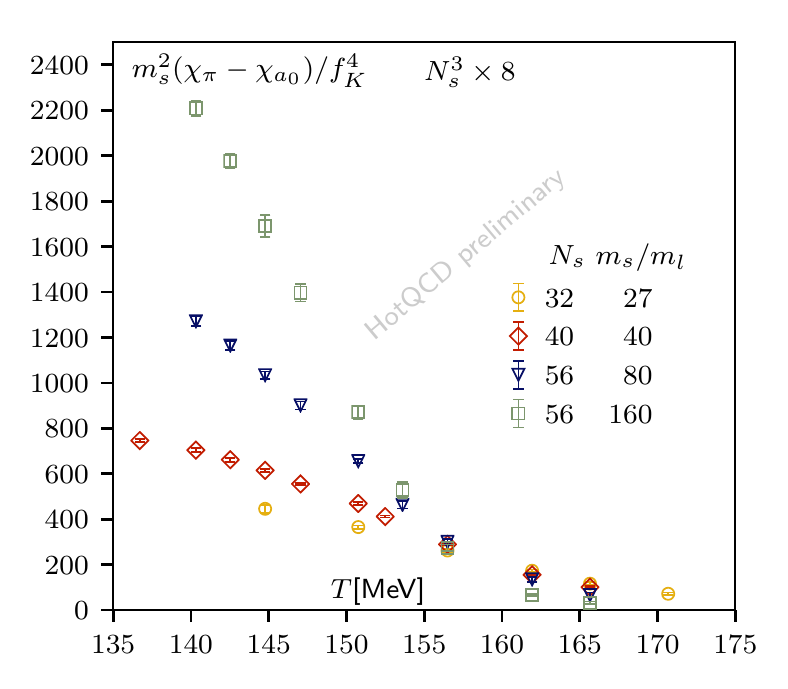} ~~~~
    \includegraphics[scale=0.45]{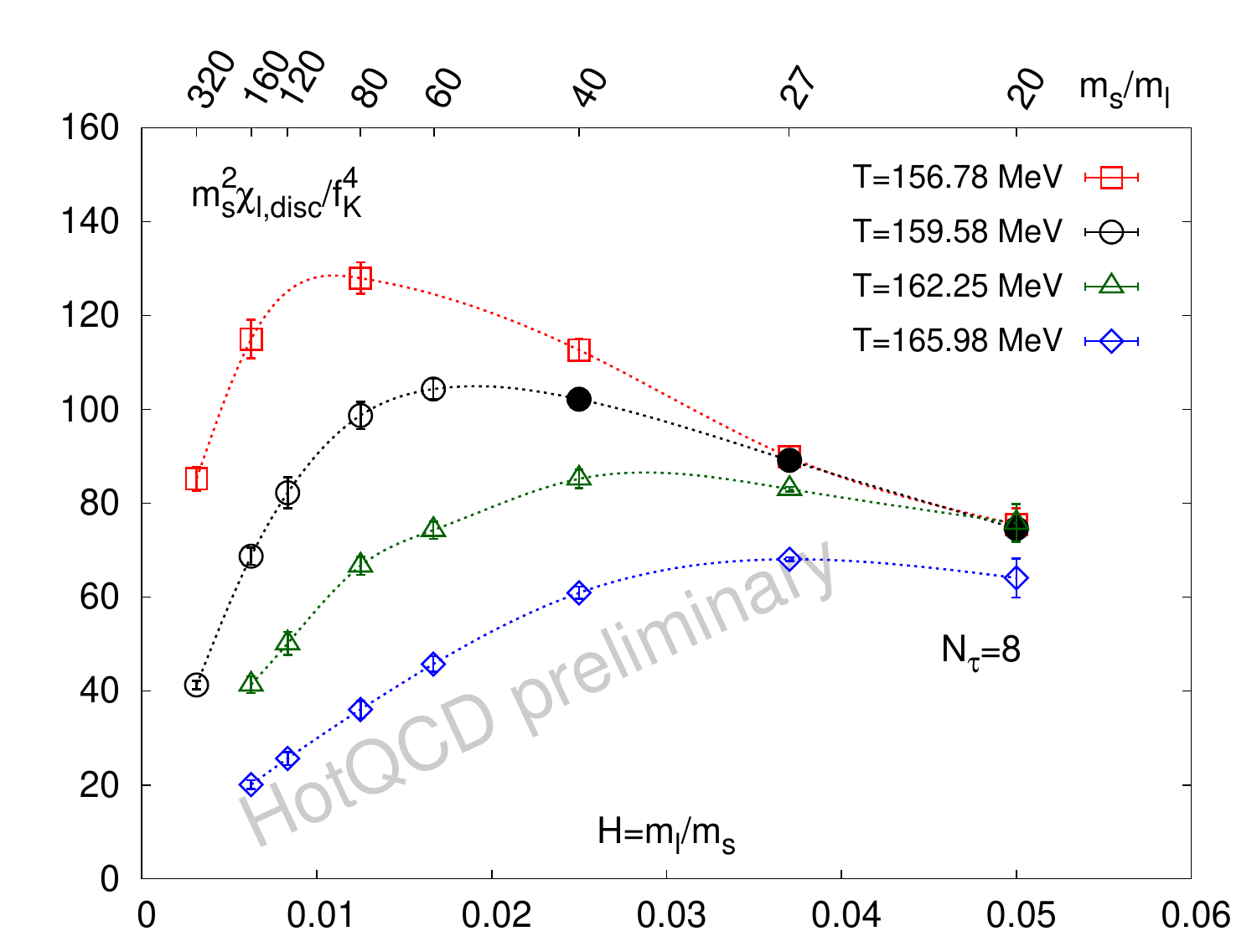}
    \caption{Left: Temperature variation of $\chi_{\pi}-\chi_{a_0}$,
    calculated using the meson correlation function using the HISQ action
    with $N_{\tau}=8$. Plot is taken from \cite{Dentinger:2021khg}.
    Right: Variation of $\chi_{\text{disc}}$ w.r.t.\ light quark mass
    for various (fixed) temperatures above but close to the chiral
    critical temperature. Calculations have been performed with the (2+1)-flavor
    HISQ action with $N_{\tau}=8$.}
    \label{fig:UA1nearTc}
\end{figure}

In the left panel of fig.\ \ref{fig:UA1nearTc} the temperature variation of $\chi_{\pi}-\chi_{a_0}$ is
shown calculated using the meson correlation function with HISQ action with $N_{\tau}=8$ \cite{Dentinger:2021khg}.
The apparent increase with smaller quark masses at low temperature is expected from the
Ward identity. At high temperatures this difference becomes small according to
the expectation, and upon looking carefully one can realize that the mass dependence
is inverted w.r.t.\ the dependence at low temperature. The exact same behavior can also be noticed
in the temperature variation of the disconnected part of the chiral susceptibility,
$\chi_{\text{disc}}$ \cite{Kaczmarek:2020sif}. It is not hard to realize that a given
fixed (intermediate) temperature simultaneously belongs to the chirally restored phase and
chirally broken phase for comparatively large and small masses, respectively.
This makes the mass dependence of any $U_A(1)$ breaking measure highly non-monotonic
for temperatures above but close to the chiral critical temperature, $T_c$.
This can be seen in a much clearer manner through the right panel of fig.\ \ref{fig:UA1nearTc}
where the mass variation of the renormalized $\chi_{\text{disc}}$ is plotted for various
temperature above but close to $T_c$ which is $\sim$ 144 MeV for $N_{\tau}=8$ \cite{HotQCD:2019xnw}.
It is clearly seen for comparatively higher temperatures, $T\gtrsim 165$ MeV,
in the right panel of fig.\ \ref{fig:UA1nearTc}. $\chi_{\text{disc}}$
monotonically decreases towards the chiral limit which is easier to handle
w.r.t.\ chiral extrapolation. On the other hand, for lower temperatures
one can clearly see the onset of the non-monotonic behavior w.r.t.\ the quark mass -
with decreasing quark masses $\chi_{\text{disc}}$ first increases to a certain
point in mass then turns and start decreasing for even lower masses.
Moreover it can be seen that this turning around towards the low values shifts
to smaller masses and the slope in quark mass becomes larger when the
temperature approaches towards $T_c$ from above which makes it really hard
to comment about the chiral limit value of $\chi_{\text{disc}}$ without having 
calculations at really small masses. Exactly same behavior has also been observed
in the study of $U_A(1)$ breaking/restoration through mesonic susceptibilities \cite{Dentinger:2021khg}.
This is one of the main difficulties in the study of effective restoration of $U_A(1)$
for temperatures above but close to $T_c$.

Interestingly the above mentioned behaviors of the $U_A(1)$ breaking measures
have a striking similarity with scaling expectation of the order parameter
susceptibility and the difference between so-called the transverse and the longitudinal
susceptibilities \cite{Engels:2011km}. The latter one has a leading quadratic
dependence on the symmetry breaking field and the former has also a quadratic
dependence but in the next-to-leading order. It can also be shown, that
this quadratic behavior can only be realized for arbitrarily small values
of the symmetry breaking field when one approaches the critical temperature
from above. At this point it is tempting to connect the discussed quadratic
behavior with the same found in ref.\ \cite{Ding:2020xlj} and could be seen in
fig.\ \ref{fig:UA1nearTc} very close to the chiral limit, although one has to admit,
it will be a bit ambiguous to associate the scaling behavior with the part(s) of
the order parameter susceptibilities which appears in the $U_A(1)$ breaking measures.
Another difficulty comes through the fact that the pseudo-critical temperatures $T_{\text{pc}}$
depend on $N_{\tau}$, meaning for a fixed mass a given temperature can simultaneously be
below and above the pseudo-critical temperature for a smaller and larger $N_{\tau}$, respectively,
which makes the continuum extrapolations very difficult.
I hope I could convince the reader why it is way more difficult to study
the breaking/restoration of $U_A(1)$ closer to $T_c$ compared to away from it.

\subsection{\boldmath$N_f=3$}

Pisarski and Wilczek originally argued that for three massless
flavors the chiral transition will be of first order, which seems
natural given that a cubic term in an effective chiral framework can be
realized for $N_f=3$ or through the critical fluctuations, even when
the strength of the anomaly is quite small near the chiral
transition temperature \cite{Pisarski:1983ms}. This finding was seconded
by a study based on renormalization-group flow where the chiral 
transition for $N_f=3$ is found to be of first order irrespective of
the effective restoration of $U_A(1)$ \cite{Pelissetto:2013hqa}.
In this regard, it appeared that the assumption of the irrelevance of
the gauge degrees of freedom in the Landau-Ginzburg-Wilson type 
approaches also should be given a serious second thought
\cite{Pelissetto:2017sfd,Pelissetto:2017pxb}. Given these apparent 
concerns it is important to ask the questions what is the order
of the chiral phase transition in QCD for three massless flavors?
Moreover, if it is first order, then what is the critical pion mass
at which the chiral transition is of second order belonging to
$Z_2$ universality class?

The community is trying to answer this question for quite a while and since
it is impossible to mention all of them in the following I try to mention
some of the recent efforts in this direction. Let me start with a
calculation with unimproved staggered actions which found a finite critical
pion mass \cite{Karsch:2001nf}, although it was realized that
the critical pion mass is very sensitive to cutoff effects. In a later
calculation with the HISQ action \cite{Bazavov:2017xul} there was no
evidence of a first order transition down to pion mass of 80 MeV, and
scaling arguments pushed the bound on critical pion mass down to 50 MeV.
Still, a possibility of a small first order region could not be
ruled out. Almost at the same time, a calculation based on
stout-smeared rooted staggered quarks \cite{Varnhorst:2015lea} observed
a significant decrease in the critical pion mass with increasing amount
of smearing, raising the possibility of having a second order phase
transition in the three flavor corner of the Columbia plot.

\begin{figure}[!h]
    \centering
    \includegraphics[scale=0.65]{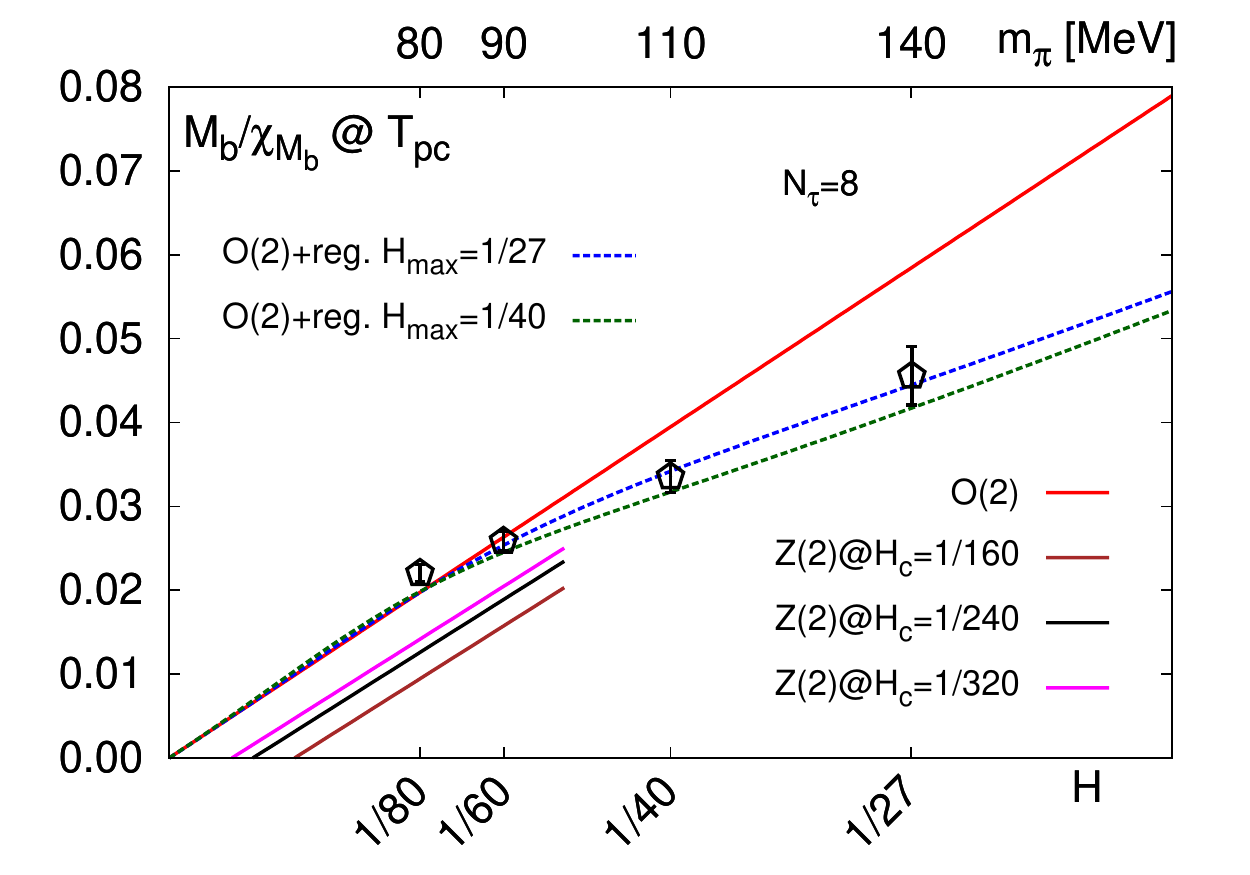} ~~~~~~~~
    \includegraphics[scale=0.5]{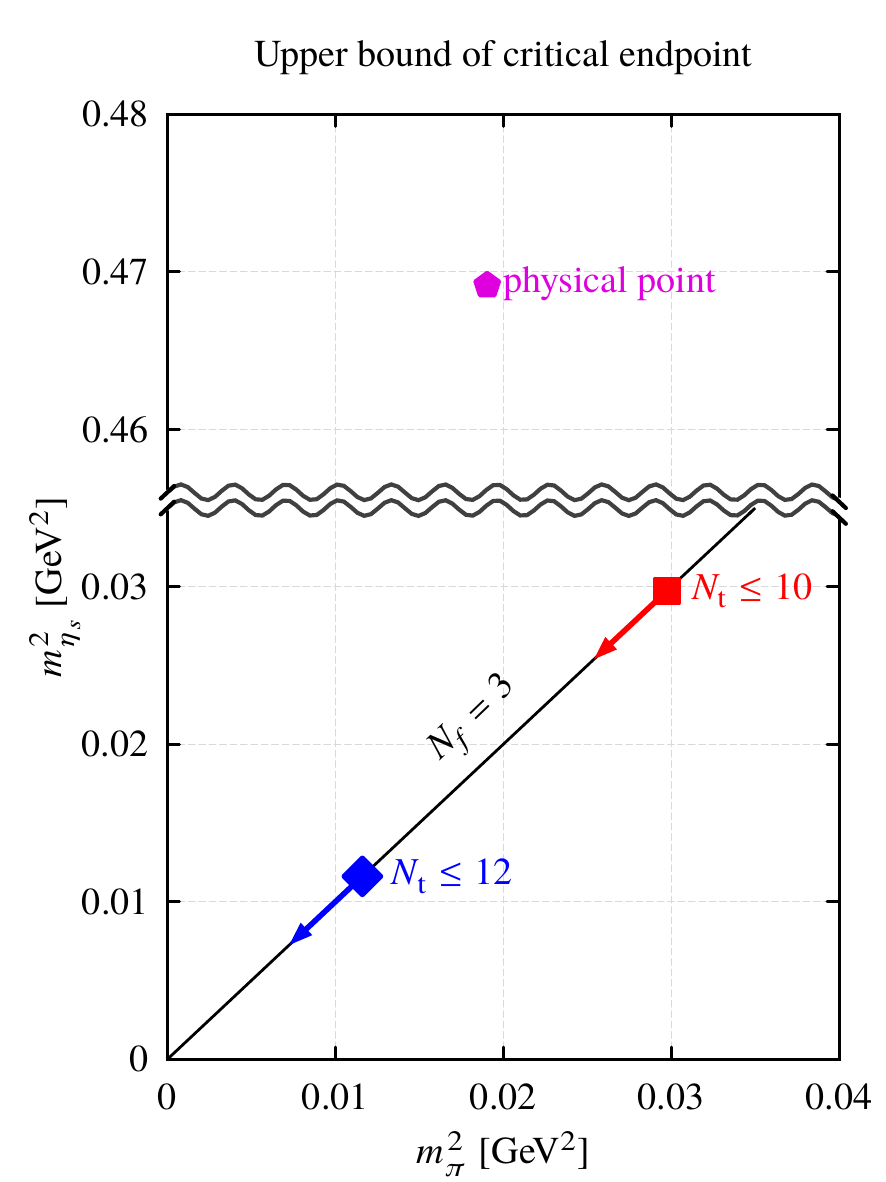}
    \caption{Left: Ratio of the non-subtracted chiral order parameter and its
    susceptibility calculated at the peak of the susceptibility shown as
    a function of mass for three degenerate flavors and compared with
    the scaling expectations. The comparison with the singular part is
    parameter free as expected from eq.\ \ref{eq:ratiounivclass}. Plot is 
    taken from ref.\ \cite{Dini:2021hug}. Right: Columbia-like plot in terms
    of hadron masses to show the critical point belonging to the $Z_2$
    universality class from calculations with ${\cal O}(a)$ improved
    Wilson fermions. The strong cutoff dependence of the critical pion mass
    is vivid. Plot is taken from ref.\ \cite{Kuramashi:2020meg}.}
    \label{fig:threeflavPTorder}
\end{figure}

Recently in a calculation using the standard staggered formulation
treating $N_f$ as a real continuous parameter \cite{Cuteri:2021ikv}, it
was found that the critical pion mass for having a second order transition
decreases with increasing $N_{\tau}$, and the scaling analysis suggests
that this critical mass eventually vanishes for finite values of
$N_{\tau}$, implying that the chiral transition for three massless
flavors will be of second order in the continuum.

In a very recent study with the HISQ action no direct evidence of a first
order phase transition was found \cite{Dini:2021hug}. Moreover it was
found that the behavior of various chiral observables made out of the
chiral condensate and its susceptibility can be very satisfactorily
described assuming a second order phase transition at the three flavor
chiral point. Further support arises from an  analysis described through 
Eq.\ \ref{eq:ratiounivclass} also for $N_f=3$. It can be clearly seen
from the left panel of fig.\ \ref{fig:threeflavPTorder} that even a
calculation with a fixed value of $N_{\tau}$ seems to prefer a
second order chiral phase transition with the correct universality class,
which is $O(2)$ in this case. The critical temperature was found to be
$T_c=98^{+3}_{-6}$ MeV, within the scale setting of (2+1)-flavor QCD.

Coming to the Wilson fermion side, the chronology has a similar 
character. In all of the calculations with ${\cal O}(a)$-improved Wilson
fermions, a finite value of critical mass is found either assuming or
establishing the $Z_2$ universality class through the calculation
of finite size scaling of the kurtosis of the chiral condensate
\cite{Jin:2014hea,Jin:2017jjp,Kuramashi:2020meg}. It is found that
although the critical temperature does not suffer much from cutoff
effects, the critical pion mass suffers from huge cutoff effects.
In the right panel of fig.\ \ref{fig:threeflavPTorder}, I borrowed a
Columbia-like plot from ref.\ \cite{Kuramashi:2020meg}, which clearly 
shows the strong cutoff dependence of the critical pion mass leaving 
only a tiny room for a first order phase transition in the chiral limit.
An interesting observation is that although of different nature, the
transition temperature found in staggered calculations \cite{Dini:2021hug}
is quite comparable with what is calculated from the Wilson studies
\cite{Jin:2017jjp,Kuramashi:2020meg}. Recently it has been shown
\cite{Cuteri:2021ikv} that the existing result for critical pion masses
from Wilson fermion calculations also show tri-critical scaling and leads to vanishing pion
masses at a finite value of $N_{\tau}$ before reaching the continuum,
implying a second order chiral phase transition.

I would like to end this part with the remark that at this point 
discussions regarding the chiral transition for $N_f=3$ are standing at a
very interesting and crucial crossing. In the following I try to list
a few comments or questions regarding the chiral transition for three
massless flavors which I collected during the discussion:
\begin{itemize}
    \item An increasing amount of evidence is coming up in favor of a
    second order chiral phase transition.
    \item What is the reason that the first order transition, expected
    for a long time, doesn't show up in QCD calculations?
    \item Related to the earlier question, what is the possible reason
    that the tri-linear coupling in the chiral effective framework originated from the anomaly contribution becomes unimportant?
    \item The origin of the observed tri-critical scaling may lie
    in a hexa-linear coupling in Landau-Ginzburg-Wilson type theories,
    as suggested in ref.\ \cite{Cuteri:2021ikv}.
    \item What is the role of gauge symmetries in the
    Landau-Ginzburg-Wilson type approaches?
\end{itemize}
Hopefully these will motivate many interesting studies in future.

\subsection{\boldmath$N_f>3$}

Theories with large number of quark flavors have also been a point of interest
of the community for various reasons. According to Pisarski and Wilczek
\cite{Pisarski:1983ms} the chiral transition is also expected to be of
first order for $N_f>3$. In the following I try to mention some recent
studies and my excuses if I miss some.

\begin{figure}[!h]
    \centering
    \includegraphics[scale=0.55]{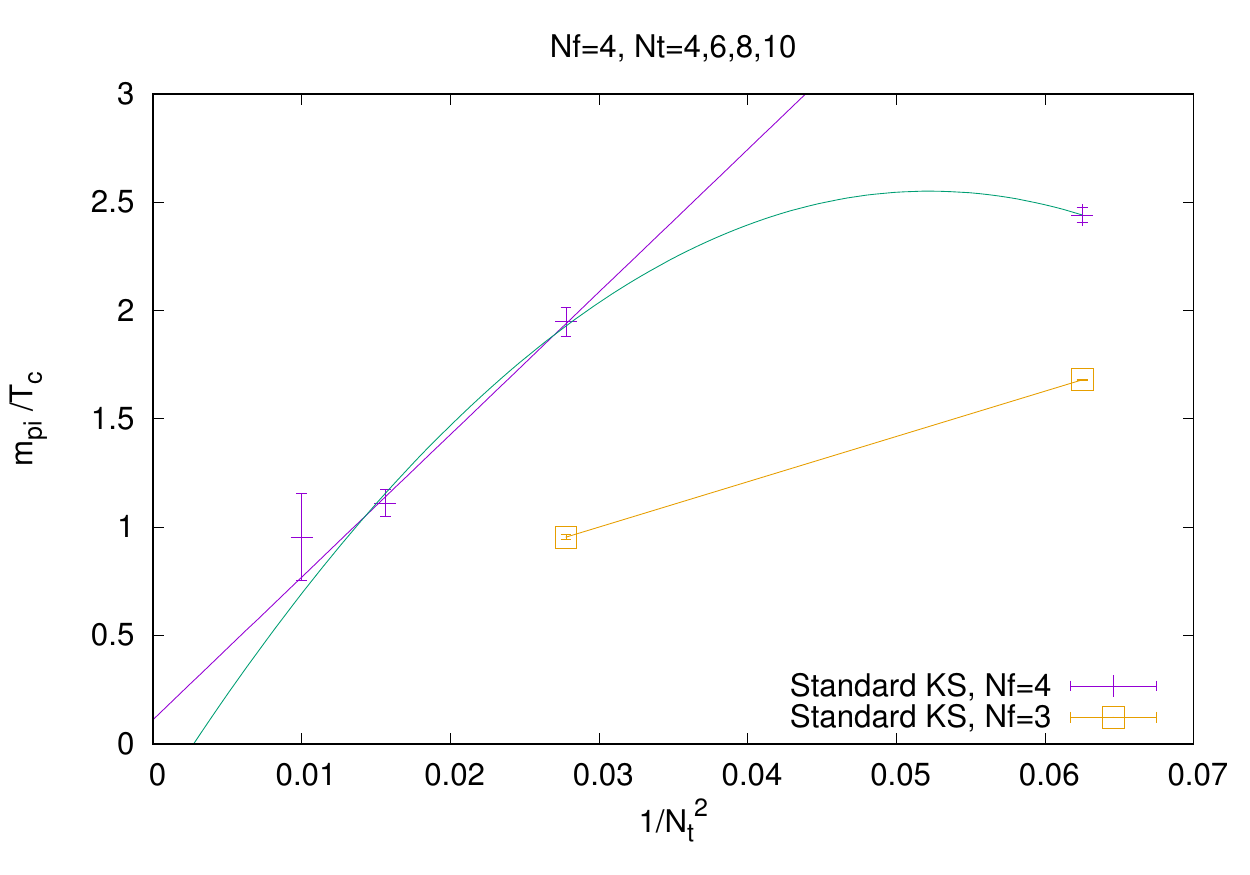} ~~~~~~
    \includegraphics[scale=0.37,angle=-90,trim={12.3cm 0 0 0}]{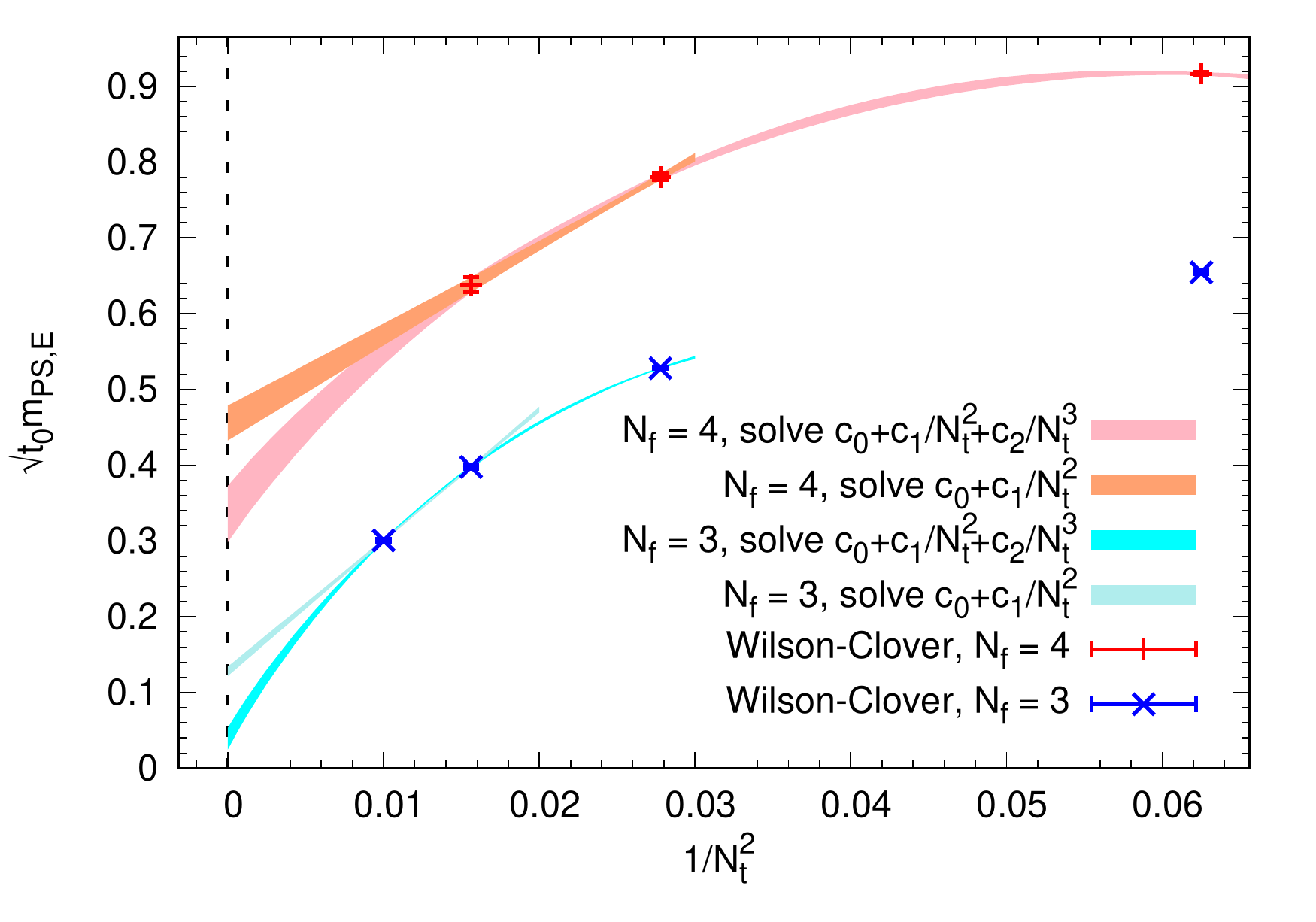}
    \caption{Left: Critical pion mass (in units of the critical 
    temperature $T_c$) for various $N_{\tau}$ from a $N_f=4$ calculation
    with unrooted standard staggered fermions. For
    comparison the critical pion mass for $N_f=3$ calculations
    are also shown. Plot is taken from  ref.\ \cite{deForcrand:2017cgb}.
    Right: Critical pion mass for various $N_{\tau}$ from a $N_f=4$
    calculation with the on-perturbatively improved Wilson-Clover
    fermion action. Plot is taken from ref.\ \cite{Ohno:2018gcx}.}
    \label{fig:mpoicrit4flav}
\end{figure}

It is usually expected that with increasing number of flavors the
first order region gets larger, which indicates to a larger value
of the critical pion mass. This expectation seems to be fulfilled
in calculations with $N_f=4$, performed with unrooted
standard staggered fermions \cite{deForcrand:2017cgb}. The strong
reduction of the critical pion mass has been realized also for $N_f=4$,
and the cutoff dependence seems to be even larger compared to $N_f=3$,
which can be seen from the left panel of fig.\ \ref{fig:mpoicrit4flav}.
This study also clarifies one important criticism about calculations
with staggered fermions that the strong reduction of the extent
of the first order region in the Columbia plot is not due to rooting. 

Coming to the calculations with Wilson fermions the situation is quite
similar to staggered one, meaning the critical pion mass for $N_f=4$
is found to be larger compared to the same for $N_f=3$ \cite{Ohno:2018gcx},
which can be seen from the right panel of fig.\ \ref{fig:mpoicrit4flav},
where apparently a non-zero critical pion mass can be seen in the
continuum limit. Since the cutoff effects are not fully under control
in ref.\ \cite{Ohno:2018gcx} an update has been planned \cite{Ohno:Lat2021} to this analysis.

Going over to even higher numbers of flavors it can be seen from the
right panel of fig.\ \ref{fig:chiralPTorder} that the cutoff effects
become larger with increasing number of flavors. It is found that
the critical pion masses calculated for $N_f=5$ also prefers
tri-critical scaling, and a clear tension can be realized while trying
to describe the critical masses with an ansatz compatible to
a first order scenario. This can only be avoided given a change
in the curvature arbitrarily close to the continuum, which would be
rather surprising. Given all these observations and realizations it
is argued that for $N_f \lesssim 6$ the first order transition can not
persist to the continuum implying that in the continuum the chiral
transition for $N_f \lesssim 6$ is going to be of second order \cite{Cuteri:2021ikv}.
This finding may have a bigger perspective w.r.t.\ to the
discussions about a conformal window of QCD when $N_f$ exceeds
some critical value \cite{Lombardo:2015oha,Braun:2009ns}.

\subsection{Energy-like observables for \boldmath$N_f=2+1$.}

For the universality classes of our interest, namely $O(N)$ or $Z_2$, in the
infinite volume limit, there exits two relevant scaling fields in the RG sense.
One, corresponding to the \textquoteleft magnetic\textquoteright-direction, breaks the
symmetry explicitly and is proportional to the symmetry breaking fields/parameters
in the leading order of Taylor expansion around the critical point.
Derivatives of the partition function w.r.t.\ this symmetry breaking field
give rise to the \textquotedblleft magnetization-like\textquotedblright\
observables {\it e.g.}\ the order parameter and its susceptibility.
The other relevant scaling field, called \textquotedblleft temperature-like\textquotedblright,\
corresponds to the \textquoteleft thermal\textquoteright-direction in RG space and
does not break the symmetry. This \textquotedblleft temperature-like\textquotedblright\
scaling field is proportional to the so-called reduced temperature to the leading order.
Important consequences arise from the realization that any parameter which does not break the
corresponding symmetry of the theory under consideration must appear in the temperature-like
scaling field. A derivative of the the partition function w.r.t.\ any such parameter,
eventually being proportional to a derivative w.r.t.\ the temperature-like scaling field,
defines an \textquotedblleft energy-like\textquotedblright\ observable. The behavior
of these observables proportional to the first and second derivatives of the partition function w.r.t.\
the temperature-like scaling field, can be related to that of the energy and specific heat in the
spin model \cite{Engels:2011km}, respectively.

Let's start with the Polyakov loop (PL), which was first proposed as an order parameter of the
confinement-deconfinement phase transition in the quenched or infinite quark mass limit of QCD  \cite{Kuti:1980gh,McLerran:1980pk,McLerran:1981pb}.
Later on there were attempts to use the inflection point of PL or the peak of the PL susceptibility
to comment about the deconfinement crossover. Some of these studies found that the crossover temperature
defined through observables derived from the chiral condensate are quite close to the same defined through
the various PL observables \cite{Cheng:2006qk,Cheng:2007jq,Bazavov:2009zn,Cheng:2009zi}, whereas others found that
the latter estimators give higher temperatures compared to the former ones \cite{Aoki:2006br,Aoki:2009sc,Bazavov:2013yv,Clarke:2019tzf}.

\begin{figure}[!h]
    \centering
    \includegraphics[scale=0.55]{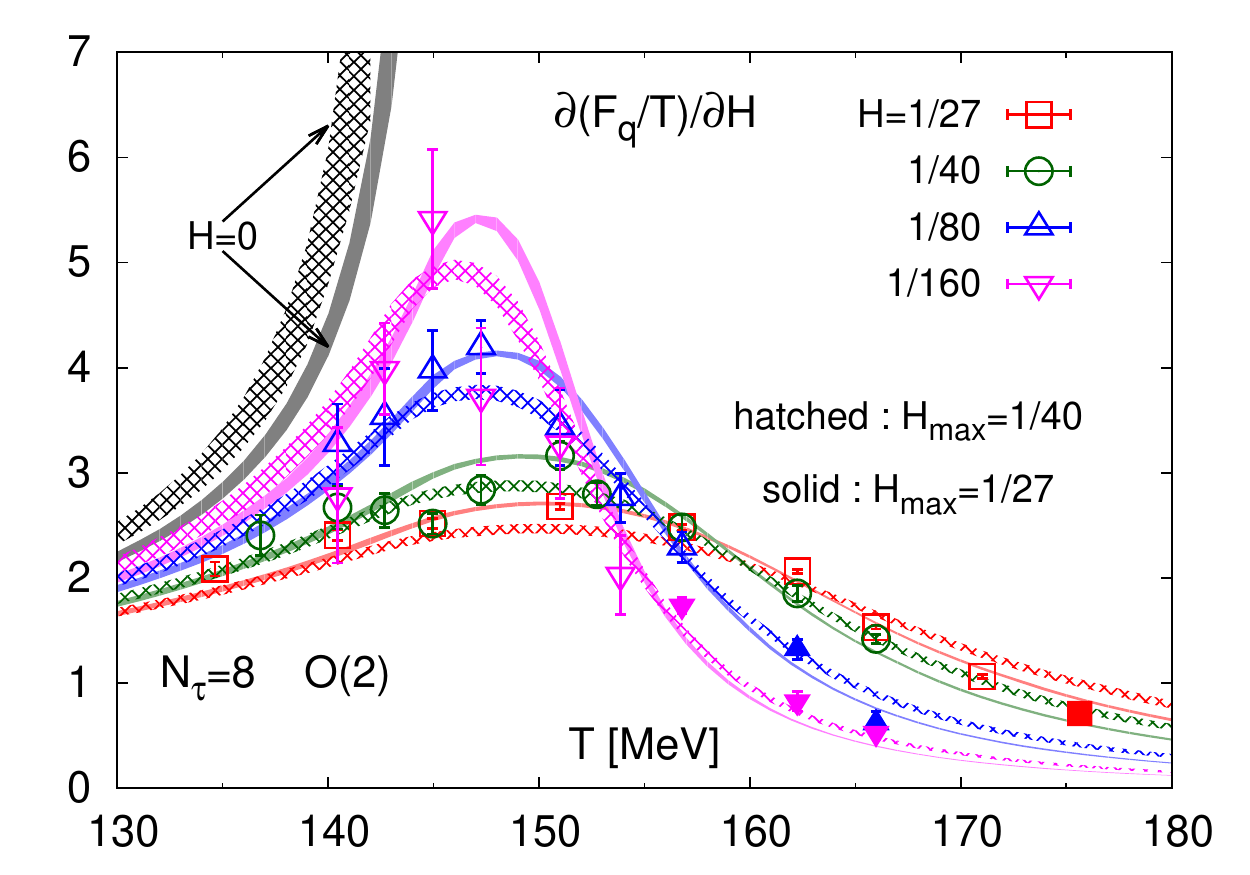} ~~~~~~
    \includegraphics[scale=0.55]{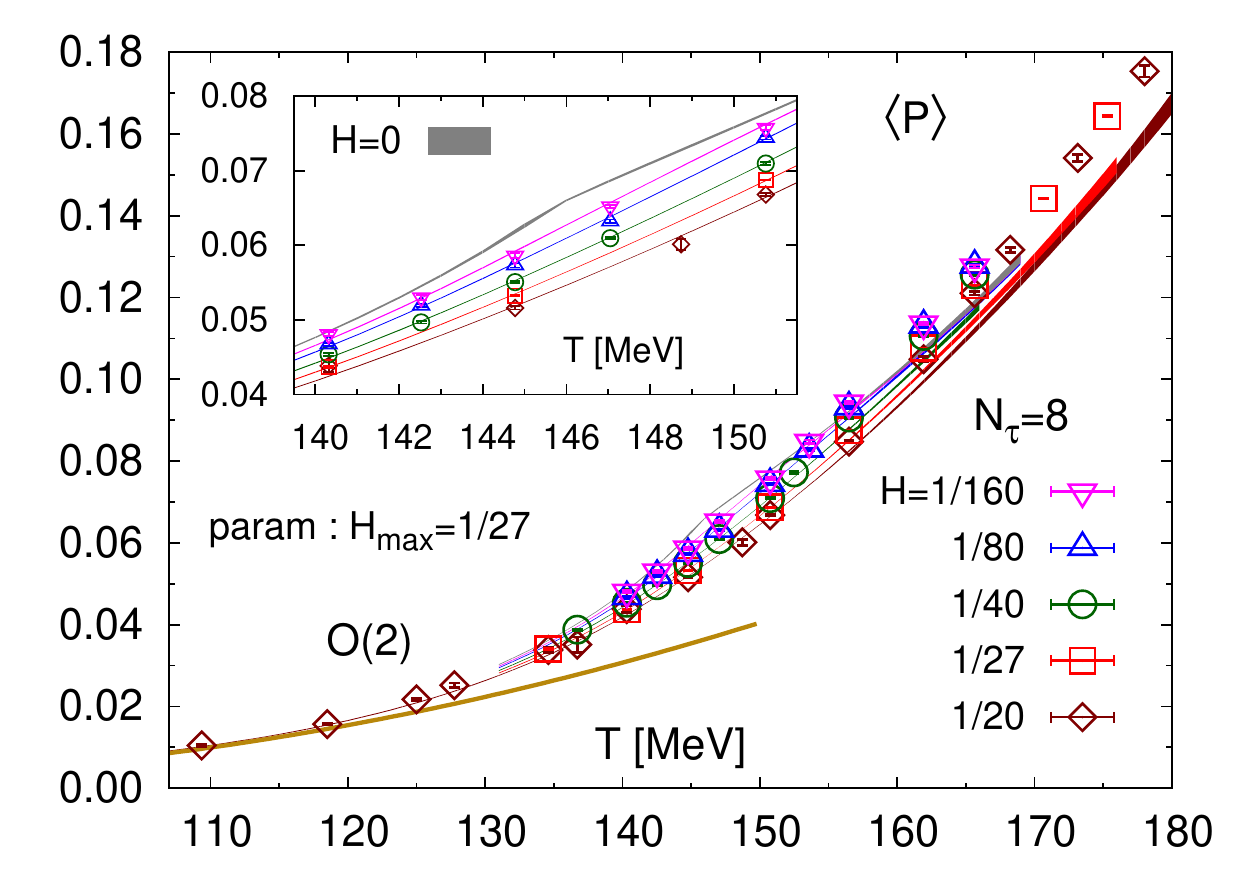}
    \caption{Left: Scaling fit to the quark mass derivative of the HQFE.
    Right: Description of the PL using the fit parameters from the scaling fits
    of the mixed-susceptibility and the HQFE.
    Calculations are performed with HISQ action with $N_{\tau}=8$.
    Both plots taken from ref.\ \cite{Clarke:2020htu}.
    }
    \label{fig:PL}
\end{figure}

Interestingly at some point it was thought that the PL is not directly sensitive to the singular structure
and none of its features are related to any critical exponent of chiral criticality \cite{Bazavov:2009zn}.
However, evolving thoughts with a gap of more than a decade
gave a completely fresh perspective of the PL w.r.t.\ the chiral phase transition \cite{Clarke:2020htu}.
Being a purely gluonic operator the PL remains invariant under global chiral rotations. Since the PL does
not appear in the QCD action directly, it will be helpful to think of an effective theory of QCD near
the critical point. In such an effective theory the PL will appear as an energy-like operator because it
does not explicitly break the chiral symmetry. Given this expectation the behavior of the PL and the related
heavy quark free energy (HQFE) are expected to behave as energy-like observables w.r.t.\ the chiral
criticality of QCD. Being energy-like, the HQFE itself is not expected to diverge at the chiral limit.
Rather a \textquoteleft mixed\textquoteright\- susceptibility, defined through the quark mass derivative
of the HQFE which is proportional to the correlation between the PL and the chiral condensate, diverges moderately 
(slower compared to the magnetic susceptibility) in the chiral limit \cite{Clarke:2020htu,Clarke:2020clx}.
In the left panel of Fig.\ \ref{fig:PL} the behavior of this mixed-susceptibility is shown towards the
chiral limit and it can be clearly seen that the mass dependence can very well be described by the
scaling expectations. Non-universal parameters extracted from the scaling fits of the mixed-susceptibility
and the regular part parameters extracted from the HQFE scaling fit describe the mass and temperature
dependence of the PL very satisfactorily, as is evident from the right panel of Fig.\ \ref{fig:PL}
ensuring the fact that the PL is indeed an energy-like observable w.r.t.\ the chiral phase transition of QCD
which calls into question treating the PL as an \textquoteleft indicator\textquoteright\ of the
deconfinement at physical and lower than physical pion masses. I shall come back later to this point
while discussing \textquotedblleft specific heat-like\textquotedblright\ observables. 

Following the discussion in the beginning of this section, one can realize that the baryon
chemical potential ($\mu_B$) does not break the chiral symmetry and it should appear in
the temperature like scaling field \cite{Sarkar:2019jwa,Sarkar:2020soa}:
\begin{equation}
    t \propto \left(\frac{T}{T_c}-1\right) + \kappa^0_B \left(\frac{\mu_B}{T_c}\right)^2
    \Rightarrow \kappa^0_B \frac{\partial}{\partial T} = \frac{T_c}{2}\frac{\partial^2}{\partial \mu_B^2},
    \text{ within the scaling window}\; ,
    \label{eq:tmuBscalingrelation}
\end{equation}
where the first term in the RHS of the proportionality relation is the usual reduced temperature and
the quadratic dependence on $\mu_B$ is to regard the ${\cal CP}$ conjugation. The non-universal parameter
$\kappa^0_B$ is the curvature of the critical line in the $T-\mu_B$ plane. This proportionality relation
implies that within the scaling window, one temperature
derivative is equivalent (with correct dimensions, of course) to a second order derivative w.r.t.\ $\mu_B$,
which immediately tells us that the second order baryon number susceptibility will be an energy-like
observable w.r.t.\ the chiral phase transition. Singular contributions to the energy-like observables,
being scaled as $H^{(1-\alpha)/\beta\delta}$ with $H$ being proportional to the light quark masses,
vanish in the chiral limit \cite{Sarkar:2019jwa,Sarkar:2020soa}.

\begin{figure}[!h]
    \centering
    \includegraphics[scale=0.55]{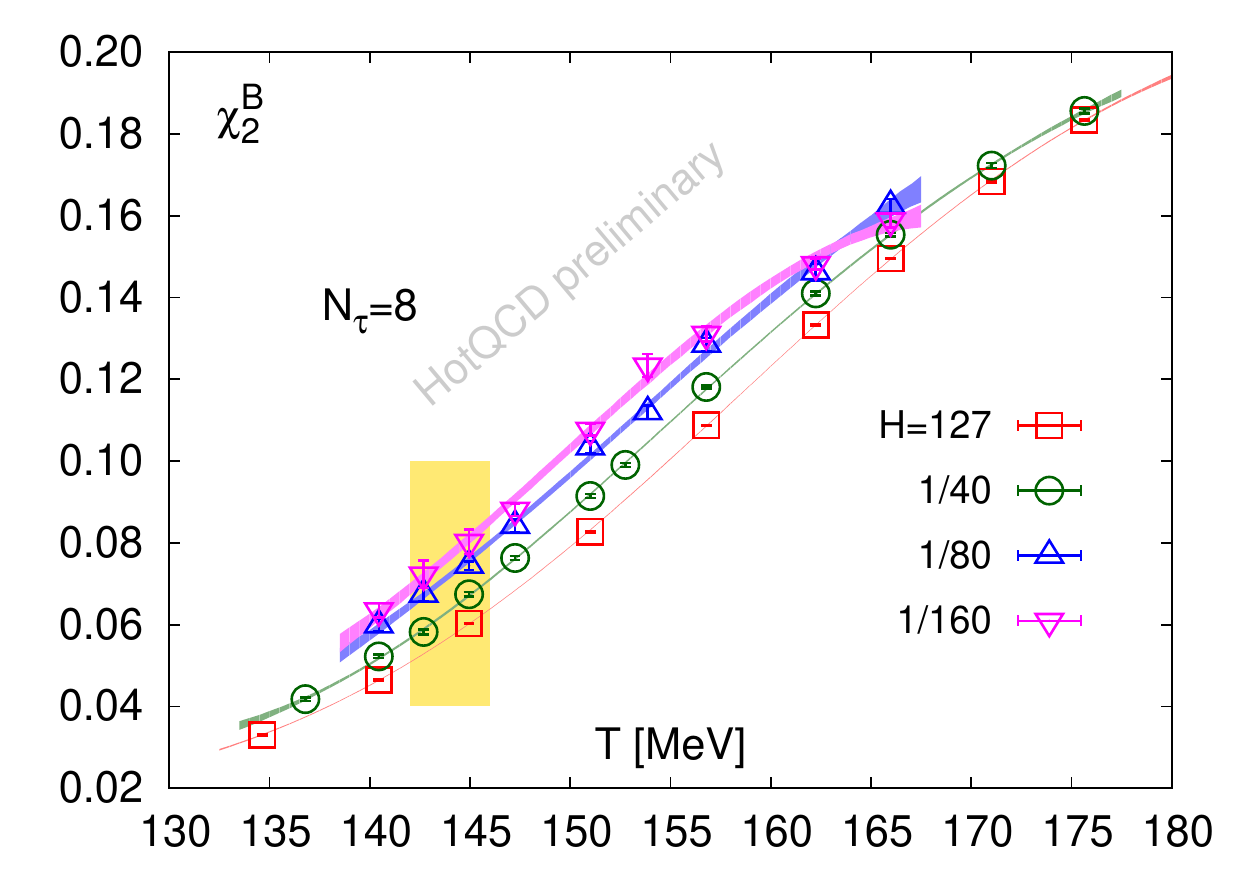} ~~~~
    \includegraphics[scale=0.55]{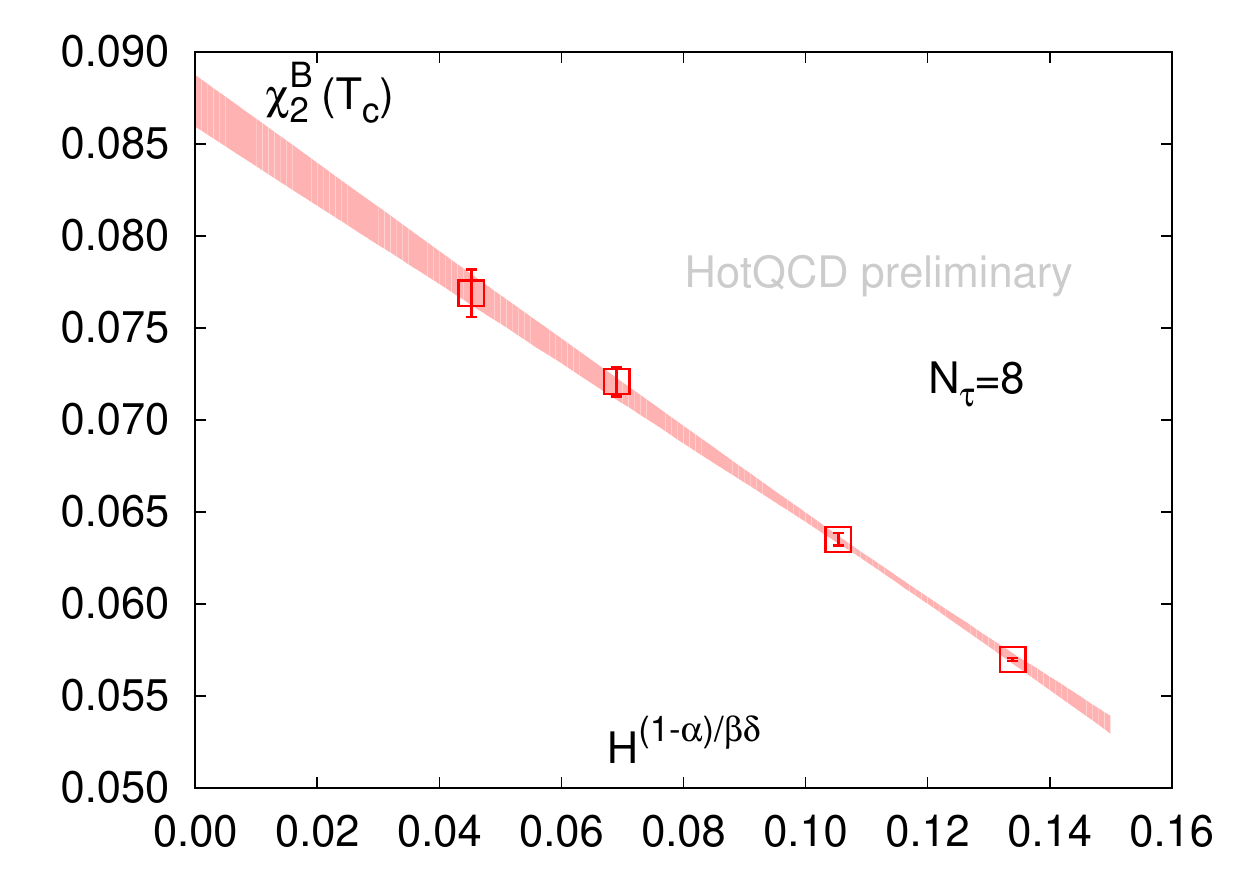}
    \caption{Left: Temperature variation of the baryon number susceptibility for various light quark masses,
    calculated with the HISQ action for $N_{\tau}=8$. The yellow band represents the chiral critical temperature,
    $T_c=144(2)$ MeV for $N_{\tau}=8$ \cite{HotQCD:2019xnw,Kaczmarek:2020err,Kaczmarek:2020sif}.
    Right: Scaling fit to the quark mass variation of the baryon number susceptibility from the left panel,
    evaluated at $T_c$, for various quark masses.}
    \label{fig:chi2B}
\end{figure}

In the left panel of Fig.\ \ref{fig:chi2B}, the baryon number susceptibility, $\chi_2^B$, for various light quark masses,
calculated with HISQ action for $N_{\tau}=8$, is shown. In the right panel of Fig.\ \ref{fig:chi2B}, it can be
seen that the scaling expectation of $\chi_2^B$ being proportional to $H^{(1-\alpha)/\beta\delta}$ is satisfied
very well. The same expectation also holds for strangeness susceptibility since the strangeness chemical potential
does not break the chiral symmetry either. Other aspects of this analysis can be found in ref.\ \cite{Sarkar:2020soa,Sarkar:2019jwa}.

Similar to chemical potentials, in (2+1)-flavor QCD, the strange quark mass $m_s$ also does not break the 2-flavor
chiral limit. Hence $m_s$ should also appear in the temperature like scaling variable and one derivative w.r.t.\ $m_s$ 
should also be equivalent (again, with proper dimensions) to one temperature derivative in the scaling regime.
As a result of this conjecture, the strange quark condensate should behave as energy-like observable w.r.t.\
the chiral transition for two massless flavors which can be ensured by calculating another mixed-susceptibility
that is essentially the correlation between condensates of light and strange quarks \cite{Sarkar:2020soa}.
Note that the fact that strange quark condensate is energy-like, may have an impact on the scaling analysis
of the subtracted chiral condensate.

Unill now I discussed some aspects of various energy-like observables. Going to the next level, observables related
to fourth order derivatives w.r.t.\ chemical potentials or one temperature and two chemical potential derivatives or
second order derivative w.r.t.\ strange quark mass of logarithm of the partition function, should behave like a specific heat,
which is obtained by two temperature derivatives of the logarithm of the partition function.
All these observables are expected to show the characteristic cusp of the $O(N)$ systems at $T_c$
whereas for small but finite quark masses they may show a rounded peak near the corresponding $T_{\text{pc}}$.
There is of course a subtlety in the form that the appearance of peaks at small masses are highly subjected to the
regular background which differs among various observables significantly. Such a background from energy density
rising with temperature was pointed out \cite{Gupta:2015dra} and confirmed \cite{HotQCD:2014kol} in a calculation
of the specific heat itself. An example can also be found from the conserved charge fluctuations where
the fourth order baryon number susceptibility shows a peak around $T_{\text{pc}}$ but
the fourth order strangeness susceptibility monotonically rises across the $T_{\text{pc}}$.
It is also difficult to locate a peak in the PL susceptibility in calculations done with improved
actions, close to the continuum limit whereas the $\mu_B$ response of the HQFE \cite{Doring:2005ih}
clearly resolves a peak already at the physical mass \cite{DElia:2019iis,Clarke:2021nrz}.
These apparently contradicting behaviors of various specific heat-like observables regarding
the peak structure needs more detailed analyses, probably focused to understanding namely whether
it is some kind of background contribution which is present in one class of observables and not in the other.

I end the discussion on the specific heat-like observables by discussing a special one - the temperature derivative of HQFE.
When calculated through temperature interpolation performed mass-by-mass, this shows a peak close to the corresponding $T_{\text{pc}}$,
and this peak  height decreases with quark mass. This behavior was proposed as a possible indicator of deconfinement
and interpreted as a hint for the possible coincidence of the chiral and deconfinement crossover \cite{Bazavov:2016uvm,Weber:2016fgn}.
On the other hand the consistent description of mass and temperature variation of the HQFE and the mixed-susceptibility by the
corresponding scaling functions, renders a scenario where peaks at finite quark masses eventually grow
(expected at quark masses which are at least two orders of magnitude smaller than what is being
used in the current day simulations) towards the
chiral limit  and show the characteristic $O(N)$ cusp at $T_c$ \cite{Clarke:2020htu}.
This asks for a serious reconsideration of the temptation to carry the interpretation of deconfinement
through the free energy or entropy of a test charge from the quenched corner to the regime of physical
and lower than physical pion masses.

I drop the scene by discussing recent efforts of calculating the curvature of the (pseudo-)critical
lines. There are mainly three avenues to calculate the curvature - using Taylor expansion,
analytic continuation of calculations done at imaginary chemical potentials or using
scaling arguments. In the former method one can choose a physical condition
at $\mu_B=0$ and follow that condition over the $T-\mu_B$ plane through the Taylor expansion
using the parametrization of the of $T_{\text{pc}}$ on $\mu_B$;
\begin{equation}
    T_{\text{pc}}\left(\mu_B\right) = T_{\text{pc}}\left(0\right)
    \left[1-\kappa^{H}_B\left(\frac{\mu_B}{T_{\text{pc}}\left(0\right)}\right)^2\right].
    \label{eq:Tpc_param}
\end{equation}
Then collecting terms in various orders in $\mu_B$ will define the curvature
within the same order of the Taylor expansion in terms of the Taylor coefficients;
{\it e.g.}\ the quadratic term will give the expression of $\kappa^{H}_B$ for this
specific value of quark mass (proportional to $H$). The choice of the physical
condition may have a wide variety, either by fixing the value of pressure, energy density
or entropy to its value at $\mu_B=0$ and $T_{\text{pc}}\left(0\right)$ \cite{Bazavov:2017dus}
or choosing a pseudo-critical condition {\it e.g.}\ the inflection point of the chiral condensate or
peak of the chiral susceptibility \cite{Bonati:2018nut,HotQCD:2018pds}. This calculation, when performed
mass-by-mass, can directly show the quark mass dependence of the curvature of the
pseudo-critical lines (represented by the black dashed line in the $T-\mu_B$ plane in fig.\ \ref{fig:3dQCDPD}).
A direct calculation of the curvature for imaginary chemical potentials using a parametrization of eq.\ \ref{eq:Tpc_param}
gives compatible results \cite{Bonati:2015bha,Borsanyi:2020fev} which rely on the validity of the analytic continuation.

\begin{wrapfigure}{l}{0.5\textwidth}
\includegraphics[scale=0.20]{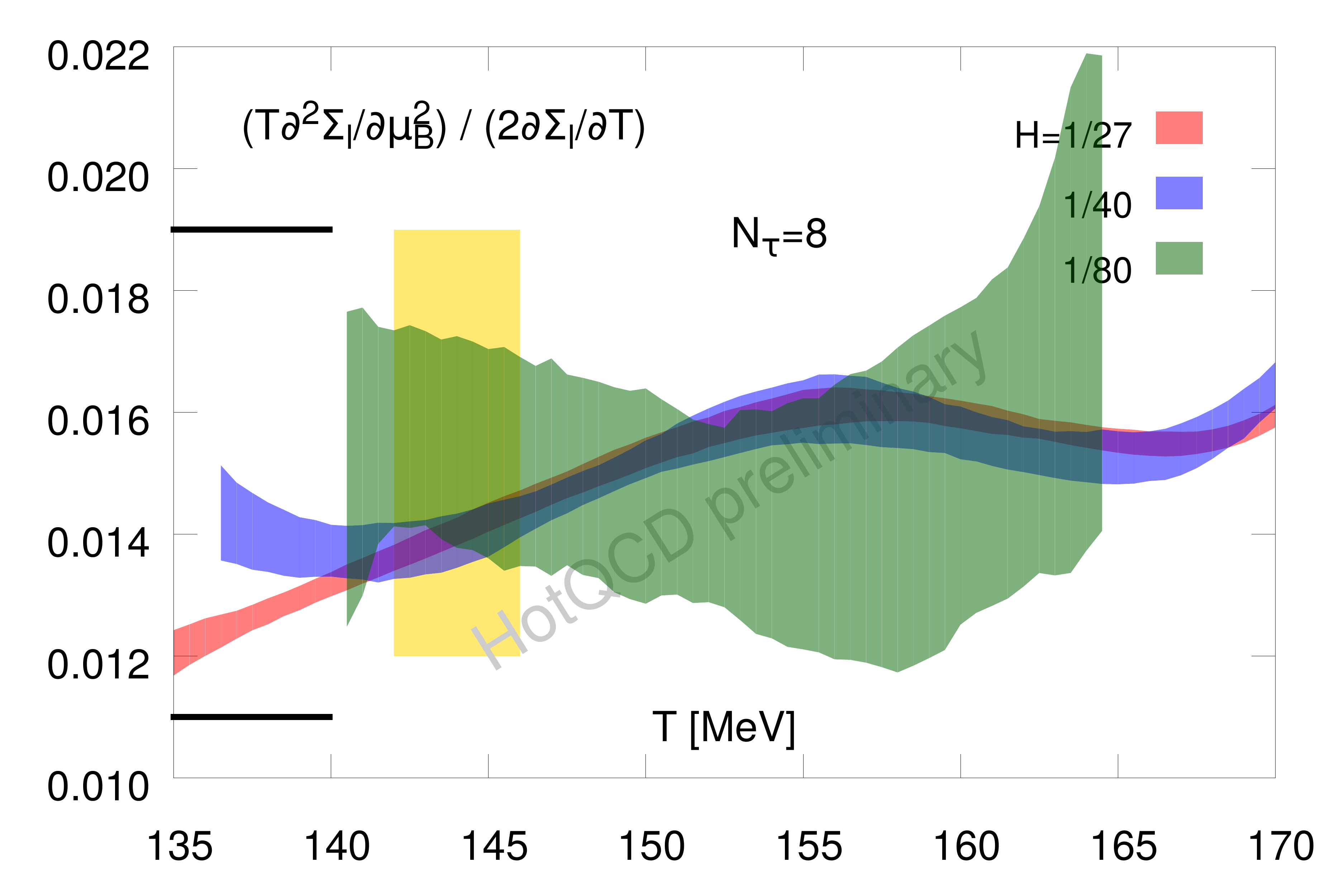}
\caption{Estimators of the curvature of the critical line in the $T-\mu_B$ plane.
Calculations are done with HISQ action for $N_{\tau}=8$. The vertical range enclosed by the pair of
black lines shows the continuum extrapolated estimate of the curvature of the pseudo-critical line
for physical pion mass \cite{HotQCD:2018pds}.}
\label{fig:curvature}
\end{wrapfigure}

The third method, which can give directly the curvature of the critical line
(represented by the red solid line in fig.\ \ref{fig:3dQCDPD}) by exploiting
the scaling relation of the mixed-susceptibility, involves the mixture of magnetic
and thermal derivatives \cite{Kaczmarek:2011zz}. There also exists a simpler way
to calculate estimators of $\kappa_B^0$ using
the implied equality in eq.\ \ref{eq:tmuBscalingrelation}; taking the ratio
between two $\mu_B$ derivative of any chiral observables and one $T$ derivative of
of the same gives an estimator of $\kappa_B^0$ for a given $H$.
In fig.\ \ref{fig:curvature} such a calculation with non-subtracted chiral condensate is shown
for various quark masses. The calculations have been done with HISQ action for $N_{\tau}=8$.
The value of the ratio around $T_c$, shown by a yellow band, is what should be looked for. In the same plot
the continuum extrapolated estimate of the curvature of the pseudo-critical line
for physical pion mass, obtained through the Taylor expansion method \cite{HotQCD:2018pds},
is indicated by the enclosed vertical interval between two black lines. This preliminary
comparison tends to suggests that the curvature may not change
dramatically towards the chiral limit \cite{Sarkar:Lat2021}.

An interesting point to note is that the ratio shown in fig.\ \ref{fig:curvature} to calculate
the estimators of $\kappa_B^0$ for a given $H$ can also be used used to estimate $\kappa_B^H$
for the same $H$ from the Taylor expansion method, corresponding to the condition that
the chiral condensate is fixed to its $\mu_B=0$ value at $T_{\text{pc}}\left(0\right)$,
over the $T-\mu_B$ plane \cite{Bazavov:2017dus}. Although then one has to focus around the pseudo-critical temperature, $T_{\text{pc}}$,
for that particular value of $H$, not around $T_c$ as is done fig.\ \ref{fig:curvature}. This suggests that
the change of $\kappa_B^H$ in the scaling regime will be $\sim$ 10\% while
going to its chiral limit value $\kappa_B^0$.

\section{Take home messages}

I hope that through this write-up of my talk I could convince the reader that study of QCD towards
the chiral limit is one of the most interesting and important fields of research.
Various groups in the lattice community have been actively working over the years
on a variety of problems and hopefully this will continue in future.
In this write-up I tried to review some of the aspects of this very dynamic field.
With this said I would like to conclude by pointing out some take home messages
in the following:

\begin{itemize}
 \item There have been some progress ....
 \begin{itemize}
  \item The chiral transition temperature for $N_f=2$ has been found to be around 130 MeV.
  \item The CEP for physical world has to be searched for $T<130$ MeV and correspondingly for $\mu_B>400$ MeV.
  \item The curvature of (pseudo-)critical lines seems to have very weak quark mass dependence towards the chiral plane.
  \item The Polyakov loop behaves as an energy-like observable w.r.t.\ the chiral phase transition, calling into
  question its relation with deconfinement, even at physical mass.
  \item Various conserved charge fluctuations and the strange condensate also behave as energy-like observables.
 \end{itemize}
 \item Feel the heat ....
 \begin{itemize}
  \item Effective restoration of $U_A(1)$ is yet to be settled among various fermion discretizations.
  \item Significantly more attention needed for temperatures higher than but close to $T_c$.
  \item The possibility of having a $1^{\text{st}}$-order region in the $N_f=3$ corner getting is feeble.
  \item Studies of many flavor QCD are going to be interesting in future, especially w.r.t.\ the
  existence of a conformal window.
 \end{itemize}
\end{itemize}

\section*{Acknowledgement}

I acknowledge support from the Deutsche Forschungsgemeinschaft (DFG, German Research
Foundation) through Project No.\ 315477589-TRR 211 and from the German Bundesministerium f\"{u}r Bildung und
Forschung through Grant No.\ 05P18PBCA1.

I would like to thank Frithjof Karsch for many enlightening discussions
and for a critical reading of this manuscript. I would also like to thank
David A.\ Clarke for a careful reading of the write-up.



\bibliographystyle{JHEP}
\bibliography{al}

\providecommand{\href}[2]{#2}\begingroup\raggedright\begin{thebibliography}{10}

\bibitem{Casher:1979vw}
A.~Casher, \emph{{Chiral Symmetry Breaking in Quark Confining Theories}},
  \href{https://doi.org/10.1016/0370-2693(79)91137-7}{\emph{Phys. Lett. B}
  {\bfseries 83} (1979) 395}.

\bibitem{Banks:1979yr}
T.~Banks and A.~Casher, \emph{{Chiral Symmetry Breaking in Confining
  Theories}}, \href{https://doi.org/10.1016/0550-3213(80)90255-2}{\emph{Nucl.
  Phys. B} {\bfseries 169} (1980) 103}.

\bibitem{Kogut:1982fn}
J.B.~Kogut, M.~Stone, H.W.~Wyld, J.~Shigemitsu, S.H.~Shenker and D.K.~Sinclair,
  \emph{{The Scales of Chiral Symmetry Breaking in Quantum Chromodynamics}},
  \href{https://doi.org/10.1103/PhysRevLett.48.1140}{\emph{Phys. Rev. Lett.}
  {\bfseries 48} (1982) 1140}.

\bibitem{Kogut:1982rt}
J.B.~Kogut, M.~Stone, H.W.~Wyld, W.R.~Gibbs, J.~Shigemitsu, S.H.~Shenker
  et~al., \emph{{Deconfinement and Chiral Symmetry Restoration at Finite
  Temperatures in SU(2) and SU(3) Gauge Theories}},
  \href{https://doi.org/10.1103/PhysRevLett.50.393}{\emph{Phys. Rev. Lett.}
  {\bfseries 50} (1983) 393}.

\bibitem{Karsch:2019mbv}
F.~Karsch, \emph{{Critical behavior and net-charge fluctuations from lattice
  QCD}}, \href{https://doi.org/10.22323/1.347.0163}{\emph{PoS} {\bfseries
  CORFU2018} (2019) 163} [\href{https://arxiv.org/abs/1905.03936}{{\ttfamily
  1905.03936}}].

\bibitem{Halasz:1998qr}
A.M.~Halasz, A.D.~Jackson, R.E.~Shrock, M.A.~Stephanov and J.J.M.~Verbaarschot,
  \emph{{On the phase diagram of QCD}},
  \href{https://doi.org/10.1103/PhysRevD.58.096007}{\emph{Phys. Rev. D}
  {\bfseries 58} (1998) 096007}
  [\href{https://arxiv.org/abs/hep-ph/9804290}{{\ttfamily hep-ph/9804290}}].

\bibitem{Pisarski:1983ms}
R.D.~Pisarski and F.~Wilczek, \emph{{Remarks on the Chiral Phase Transition in
  Chromodynamics}}, \href{https://doi.org/10.1103/PhysRevD.29.338}{\emph{Phys.
  Rev. D} {\bfseries 29} (1984) 338}.

\bibitem{Hatta:2002sj}
Y.~Hatta and T.~Ikeda, \emph{{Universality, the QCD critical / tricritical
  point and the quark number susceptibility}},
  \href{https://doi.org/10.1103/PhysRevD.67.014028}{\emph{Phys. Rev. D}
  {\bfseries 67} (2003) 014028}
  [\href{https://arxiv.org/abs/hep-ph/0210284}{{\ttfamily hep-ph/0210284}}].

\bibitem{HotQCD:2018pds}
{\scshape HotQCD} collaboration, \emph{{Chiral crossover in QCD at zero and
  non-zero chemical potentials}},
  \href{https://doi.org/10.1016/j.physletb.2019.05.013}{\emph{Phys. Lett. B}
  {\bfseries 795} (2019) 15}
  [\href{https://arxiv.org/abs/1812.08235}{{\ttfamily 1812.08235}}].

\bibitem{Borsanyi:2020fev}
S.~Borsanyi, Z.~Fodor, J.N.~Guenther, R.~Kara, S.D.~Katz, P.~Parotto et~al.,
  \emph{{QCD Crossover at Finite Chemical Potential from Lattice Simulations}},
  \href{https://doi.org/10.1103/PhysRevLett.125.052001}{\emph{Phys. Rev. Lett.}
  {\bfseries 125} (2020) 052001}
  [\href{https://arxiv.org/abs/2002.02821}{{\ttfamily 2002.02821}}].

\bibitem{Pelissetto:2013hqa}
A.~Pelissetto and E.~Vicari, \emph{{Relevance of the axial anomaly at the
  finite-temperature chiral transition in QCD}},
  \href{https://doi.org/10.1103/PhysRevD.88.105018}{\emph{Phys. Rev. D}
  {\bfseries 88} (2013) 105018}
  [\href{https://arxiv.org/abs/1309.5446}{{\ttfamily 1309.5446}}].

\bibitem{Philipsen:2016hkv}
O.~Philipsen and C.~Pinke, \emph{{The $N_f=2$ QCD chiral phase transition with
  Wilson fermions at zero and imaginary chemical potential}},
  \href{https://doi.org/10.1103/PhysRevD.93.114507}{\emph{Phys. Rev. D}
  {\bfseries 93} (2016) 114507}
  [\href{https://arxiv.org/abs/1602.06129}{{\ttfamily 1602.06129}}].

\bibitem{Brown:1990ev}
F.R.~Brown, F.P.~Butler, H.~Chen, N.H.~Christ, Z.-h.~Dong, W.~Schaffer et~al.,
  \emph{{On the existence of a phase transition for QCD with three light
  quarks}}, \href{https://doi.org/10.1103/PhysRevLett.65.2491}{\emph{Phys. Rev.
  Lett.} {\bfseries 65} (1990) 2491}.

\bibitem{Gupta:2008ac}
S.~Gupta, \emph{{Phases and properties of quark matter}},
  \href{https://doi.org/10.1088/0954-3899/35/10/104018}{\emph{J. Phys. G}
  {\bfseries 35} (2008) 104018}
  [\href{https://arxiv.org/abs/0806.2255}{{\ttfamily 0806.2255}}].

\bibitem{HotQCD:2019xnw}
{\scshape HotQCD} collaboration, \emph{{Chiral Phase Transition Temperature in
  ( 2+1 )-Flavor QCD}},
  \href{https://doi.org/10.1103/PhysRevLett.123.062002}{\emph{Phys. Rev. Lett.}
  {\bfseries 123} (2019) 062002}
  [\href{https://arxiv.org/abs/1903.04801}{{\ttfamily 1903.04801}}].

\bibitem{Kaczmarek:2020err}
O.~Kaczmarek, F.~Karsch, A.~Lahiri and C.~Schmidt, \emph{{Universal scaling
  properties of QCD close to the chiral limit}},
  \href{https://doi.org/10.5506/APHYSPOLBSUPP.14.291}{\emph{Acta Phys. Polon.
  Supp.} {\bfseries 14} (2021) 291}
  [\href{https://arxiv.org/abs/2010.15593}{{\ttfamily 2010.15593}}].

\bibitem{Kotov:2021rah}
A.Y.~Kotov, M.P.~Lombardo and A.~Trunin, \emph{{QCD transition at the physical
  point, and its scaling window from twisted mass Wilson fermions}},
  \href{https://doi.org/10.1016/j.physletb.2021.136749}{\emph{Phys. Lett. B}
  {\bfseries 823} (2021) 136749}
  [\href{https://arxiv.org/abs/2105.09842}{{\ttfamily 2105.09842}}].

\bibitem{Cuteri:2021ikv}
F.~Cuteri, O.~Philipsen and A.~Sciarra, \emph{{On the order of the QCD chiral
  phase transition for different numbers of quark flavours}},
  \href{https://doi.org/10.1007/JHEP11(2021)141}{\emph{JHEP} {\bfseries 11}
  (2021) 141} [\href{https://arxiv.org/abs/2107.12739}{{\ttfamily
  2107.12739}}].

\bibitem{HotQCD:2012vvd}
{\scshape HotQCD} collaboration, \emph{{The chiral transition and $U(1)_A$
  symmetry restoration from lattice QCD using Domain Wall Fermions}},
  \href{https://doi.org/10.1103/PhysRevD.86.094503}{\emph{Phys. Rev. D}
  {\bfseries 86} (2012) 094503}
  [\href{https://arxiv.org/abs/1205.3535}{{\ttfamily 1205.3535}}].

\bibitem{Kaczmarek:2020sif}
O.~Kaczmarek, F.~Karsch, A.~Lahiri, L.~Mazur and C.~Schmidt, \emph{{QCD phase
  transition in the chiral limit}},  3, 2020
  [\href{https://arxiv.org/abs/2003.07920}{{\ttfamily 2003.07920}}].

\bibitem{Buchoff:2013nra}
M.I.~Buchoff et~al., \emph{{QCD chiral transition, U(1)A symmetry and the dirac
  spectrum using domain wall fermions}},
  \href{https://doi.org/10.1103/PhysRevD.89.054514}{\emph{Phys. Rev. D}
  {\bfseries 89} (2014) 054514}
  [\href{https://arxiv.org/abs/1309.4149}{{\ttfamily 1309.4149}}].

\bibitem{Bhattacharya:2014ara}
T.~Bhattacharya et~al., \emph{{QCD Phase Transition with Chiral Quarks and
  Physical Quark Masses}},
  \href{https://doi.org/10.1103/PhysRevLett.113.082001}{\emph{Phys. Rev. Lett.}
  {\bfseries 113} (2014) 082001}
  [\href{https://arxiv.org/abs/1402.5175}{{\ttfamily 1402.5175}}].

\bibitem{Cossu:2013uua}
G.~Cossu, S.~Aoki, H.~Fukaya, S.~Hashimoto, T.~Kaneko, H.~Matsufuru et~al.,
  \emph{{Finite temperature study of the axial U(1) symmetry on the lattice
  with overlap fermion formulation}},
  \href{https://doi.org/10.1103/PhysRevD.87.114514}{\emph{Phys. Rev. D}
  {\bfseries 87} (2013) 114514}
  [\href{https://arxiv.org/abs/1304.6145}{{\ttfamily 1304.6145}}].

\bibitem{Dick:2015twa}
V.~Dick, F.~Karsch, E.~Laermann, S.~Mukherjee and S.~Sharma, \emph{{Microscopic
  origin of $U_A(1)$ symmetry violation in the high temperature phase of QCD}},
  \href{https://doi.org/10.1103/PhysRevD.91.094504}{\emph{Phys. Rev. D}
  {\bfseries 91} (2015) 094504}
  [\href{https://arxiv.org/abs/1502.06190}{{\ttfamily 1502.06190}}].

\bibitem{Brandt:2016daq}
B.B.~Brandt, A.~Francis, H.B.~Meyer, O.~Philipsen, D.~Robaina and H.~Wittig,
  \emph{{On the strength of the $U_A(1)$ anomaly at the chiral phase transition
  in $N_f=2$ QCD}}, \href{https://doi.org/10.1007/JHEP12(2016)158}{\emph{JHEP}
  {\bfseries 12} (2016) 158}
  [\href{https://arxiv.org/abs/1608.06882}{{\ttfamily 1608.06882}}].

\bibitem{Ding:2020xlj}
H.T.~Ding, S.T.~Li, S.~Mukherjee, A.~Tomiya, X.D.~Wang and Y.~Zhang,
  \emph{{Correlated Dirac Eigenvalues and Axial Anomaly in Chiral Symmetric
  QCD}}, \href{https://doi.org/10.1103/PhysRevLett.126.082001}{\emph{Phys. Rev.
  Lett.} {\bfseries 126} (2021) 082001}
  [\href{https://arxiv.org/abs/2010.14836}{{\ttfamily 2010.14836}}].

\bibitem{Ding:2021gdy}
H.-T.~Ding, W.-P.~Huang, M.~Lin, S.~Mukherjee, P.~Petreczky and Y.~Zhang,
  \emph{{Correlated Dirac eigenvalues around the transition temperature on
  $N_{\tau}=8$ lattices}},  in \emph{{38th International Symposium on Lattice
  Field Theory}}, 12, 2021 [\href{https://arxiv.org/abs/2112.00318}{{\ttfamily
  2112.00318}}].

\bibitem{Tomiya:2016jwr}
A.~Tomiya, G.~Cossu, S.~Aoki, H.~Fukaya, S.~Hashimoto, T.~Kaneko et~al.,
  \emph{{Evidence of effective axial U(1) symmetry restoration at high
  temperature QCD}},
  \href{https://doi.org/10.1103/PhysRevD.96.034509}{\emph{Phys. Rev. D}
  {\bfseries 96} (2017) 034509}
  [\href{https://arxiv.org/abs/1612.01908}{{\ttfamily 1612.01908}}].

\bibitem{Kaczmarek:2021ser}
O.~Kaczmarek, L.~Mazur and S.~Sharma, \emph{{Eigenvalue spectra of QCD and the
  fate of UA(1) breaking towards the chiral limit}},
  \href{https://doi.org/10.1103/PhysRevD.104.094518}{\emph{Phys. Rev. D}
  {\bfseries 104} (2021) 094518}
  [\href{https://arxiv.org/abs/2102.06136}{{\ttfamily 2102.06136}}].

\bibitem{Aoki:2021qws}
{\scshape JLQCD} collaboration, \emph{{Role of axial U(1) anomaly in chiral
  susceptibility of QCD at high temperature}},
  \href{https://arxiv.org/abs/2103.05954}{{\ttfamily 2103.05954}}.

\bibitem{Dentinger:2021khg}
S.~Dentinger, O.~Kaczmarek and A.~Lahiri, \emph{{Screening masses towards
  chiral limit}},
  \href{https://doi.org/10.5506/APHYSPOLBSUPP.14.321}{\emph{Acta Phys. Polon.
  Supp.} {\bfseries 14} (2021) 321}
  [\href{https://arxiv.org/abs/2102.09916}{{\ttfamily 2102.09916}}].

\bibitem{Engels:2011km}
J.~Engels and F.~Karsch, \emph{{The scaling functions of the free energy
  density and its derivatives for the 3d O(4) model}},
  \href{https://doi.org/10.1103/PhysRevD.85.094506}{\emph{Phys. Rev. D}
  {\bfseries 85} (2012) 094506}
  [\href{https://arxiv.org/abs/1105.0584}{{\ttfamily 1105.0584}}].

\bibitem{Pelissetto:2017sfd}
A.~Pelissetto, A.~Tripodo and E.~Vicari, \emph{{Landau-Ginzburg-Wilson approach
  to critical phenomena in the presence of gauge symmetries}},
  \href{https://doi.org/10.1103/PhysRevD.96.034505}{\emph{Phys. Rev. D}
  {\bfseries 96} (2017) 034505}
  [\href{https://arxiv.org/abs/1706.04365}{{\ttfamily 1706.04365}}].

\bibitem{Pelissetto:2017pxb}
A.~Pelissetto, A.~Tripodo and E.~Vicari, \emph{{Criticality of O(N) symmetric
  models in the presence of discrete gauge symmetries}},
  \href{https://doi.org/10.1103/PhysRevE.97.012123}{\emph{Phys. Rev. E}
  {\bfseries 97} (2018) 012123}
  [\href{https://arxiv.org/abs/1711.04567}{{\ttfamily 1711.04567}}].

\bibitem{Karsch:2001nf}
F.~Karsch, E.~Laermann and C.~Schmidt, \emph{{The Chiral critical point in
  three-flavor QCD}},
  \href{https://doi.org/10.1016/S0370-2693(01)01114-5}{\emph{Phys. Lett. B}
  {\bfseries 520} (2001) 41}
  [\href{https://arxiv.org/abs/hep-lat/0107020}{{\ttfamily hep-lat/0107020}}].

\bibitem{Bazavov:2017xul}
A.~Bazavov, H.T.~Ding, P.~Hegde, F.~Karsch, E.~Laermann, S.~Mukherjee et~al.,
  \emph{{Chiral phase structure of three flavor QCD at vanishing baryon number
  density}}, \href{https://doi.org/10.1103/PhysRevD.95.074505}{\emph{Phys. Rev.
  D} {\bfseries 95} (2017) 074505}
  [\href{https://arxiv.org/abs/1701.03548}{{\ttfamily 1701.03548}}].

\bibitem{Varnhorst:2015lea}
L.~Varnhorst, \emph{{The $N_f$=3 critical endpoint with smeared staggered
  quarks}}, \href{https://doi.org/10.22323/1.214.0193}{\emph{PoS} {\bfseries
  LATTICE2014} (2015) 193}.

\bibitem{Dini:2021hug}
L.~Dini, P.~Hegde, F.~Karsch, A.~Lahiri, C.~Schmidt and S.~Sharma, \emph{{The
  Chiral Phase Transition in 3-flavor QCD from Lattice QCD}},
  \href{https://arxiv.org/abs/2111.12599}{{\ttfamily 2111.12599}}.

\bibitem{Kuramashi:2020meg}
Y.~Kuramashi, Y.~Nakamura, H.~Ohno and S.~Takeda, \emph{{Nature of the phase
  transition for finite temperature $N_{\rm f}=3$ QCD with nonperturbatively
  O($a$) improved Wilson fermions at $N_{\rm t}=12$}},
  \href{https://doi.org/10.1103/PhysRevD.101.054509}{\emph{Phys. Rev. D}
  {\bfseries 101} (2020) 054509}
  [\href{https://arxiv.org/abs/2001.04398}{{\ttfamily 2001.04398}}].

\bibitem{Jin:2014hea}
X.-Y.~Jin, Y.~Kuramashi, Y.~Nakamura, S.~Takeda and A.~Ukawa, \emph{{Critical
  endpoint of the finite temperature phase transition for three flavor QCD}},
  \href{https://doi.org/10.1103/PhysRevD.91.014508}{\emph{Phys. Rev. D}
  {\bfseries 91} (2015) 014508}
  [\href{https://arxiv.org/abs/1411.7461}{{\ttfamily 1411.7461}}].

\bibitem{Jin:2017jjp}
X.-Y.~Jin, Y.~Kuramashi, Y.~Nakamura, S.~Takeda and A.~Ukawa, \emph{{Critical
  point phase transition for finite temperature 3-flavor QCD with
  non-perturbatively O($a$) improved Wilson fermions at $N_{\rm t}=10$}},
  \href{https://doi.org/10.1103/PhysRevD.96.034523}{\emph{Phys. Rev. D}
  {\bfseries 96} (2017) 034523}
  [\href{https://arxiv.org/abs/1706.01178}{{\ttfamily 1706.01178}}].

\bibitem{deForcrand:2017cgb}
P.~de~Forcrand and M.~D'Elia, \emph{{Continuum limit and universality of the
  Columbia plot}}, \href{https://doi.org/10.22323/1.256.0081}{\emph{PoS}
  {\bfseries LATTICE2016} (2017) 081}
  [\href{https://arxiv.org/abs/1702.00330}{{\ttfamily 1702.00330}}].

\bibitem{Ohno:2018gcx}
H.~Ohno, Y.~Kuramashi, Y.~Nakamura and S.~Takeda, \emph{{Continuum
  extrapolation of the critical endpoint in 4-flavor QCD with Wilson-Clover
  fermions}}, \href{https://doi.org/10.22323/1.334.0174}{\emph{PoS} {\bfseries
  LATTICE2018} (2018) 174} [\href{https://arxiv.org/abs/1812.01318}{{\ttfamily
  1812.01318}}].

\bibitem{Ohno:Lat2021}
H.~Ohno, ``Critical endpoints in (2+1)-and 4-flavor qcd with wilson-clover
  fermions - in this conference.''.

\bibitem{Lombardo:2015oha}
M.P.~Lombardo, K.~Miura, T.J.~Nunes~da Silva and E.~Pallante, \emph{{Chiral
  symmetry restoration in QCD with many flavors}},
  \href{https://doi.org/10.22323/1.217.0059}{\emph{PoS} {\bfseries CPOD2014}
  (2015) 059} [\href{https://arxiv.org/abs/1506.05946}{{\ttfamily
  1506.05946}}].

\bibitem{Braun:2009ns}
J.~Braun and H.~Gies, \emph{{Scaling laws near the conformal window of
  many-flavor QCD}}, \href{https://doi.org/10.1007/JHEP05(2010)060}{\emph{JHEP}
  {\bfseries 05} (2010) 060} [\href{https://arxiv.org/abs/0912.4168}{{\ttfamily
  0912.4168}}].

\bibitem{Kuti:1980gh}
J.~Kuti, J.~Polonyi and K.~Szlachanyi, \emph{{Monte Carlo Study of SU(2) Gauge
  Theory at Finite Temperature}},
  \href{https://doi.org/10.1016/0370-2693(81)90987-4}{\emph{Phys. Lett. B}
  {\bfseries 98} (1981) 199}.

\bibitem{McLerran:1980pk}
L.D.~McLerran and B.~Svetitsky, \emph{{A Monte Carlo Study of SU(2) Yang-Mills
  Theory at Finite Temperature}},
  \href{https://doi.org/10.1016/0370-2693(81)90986-2}{\emph{Phys. Lett. B}
  {\bfseries 98} (1981) 195}.

\bibitem{McLerran:1981pb}
L.D.~McLerran and B.~Svetitsky, \emph{Quark liberation at high temperature: A
  monte carlo study of {SU(2)} gauge theory},
  \href{https://doi.org/10.1103/PhysRevD.24.450}{\emph{Phys.\ Rev.\ D}
  {\bfseries 24} (1981) 450}.

\bibitem{Cheng:2006qk}
M.~Cheng et~al., \emph{{The Transition temperature in QCD}},
  \href{https://doi.org/10.1103/PhysRevD.74.054507}{\emph{Phys. Rev. D}
  {\bfseries 74} (2006) 054507}
  [\href{https://arxiv.org/abs/hep-lat/0608013}{{\ttfamily hep-lat/0608013}}].

\bibitem{Cheng:2007jq}
M.~Cheng et~al., \emph{{The QCD equation of state with almost physical quark
  masses}}, \href{https://doi.org/10.1103/PhysRevD.77.014511}{\emph{Phys. Rev.
  D} {\bfseries 77} (2008) 014511}
  [\href{https://arxiv.org/abs/0710.0354}{{\ttfamily 0710.0354}}].

\bibitem{Bazavov:2009zn}
A.~Bazavov et~al., \emph{{Equation of state and QCD transition at finite
  temperature}}, \href{https://doi.org/10.1103/PhysRevD.80.014504}{\emph{Phys.
  Rev. D} {\bfseries 80} (2009) 014504}
  [\href{https://arxiv.org/abs/0903.4379}{{\ttfamily 0903.4379}}].

\bibitem{Cheng:2009zi}
M.~Cheng et~al., \emph{{Equation of State for physical quark masses}},
  \href{https://doi.org/10.1103/PhysRevD.81.054504}{\emph{Phys. Rev. D}
  {\bfseries 81} (2010) 054504}
  [\href{https://arxiv.org/abs/0911.2215}{{\ttfamily 0911.2215}}].

\bibitem{Aoki:2006br}
Y.~Aoki, Z.~Fodor, S.D.~Katz and K.K.~Szabo, \emph{{The QCD transition
  temperature: Results with physical masses in the continuum limit}},
  \href{https://doi.org/10.1016/j.physletb.2006.10.021}{\emph{Phys. Lett. B}
  {\bfseries 643} (2006) 46}
  [\href{https://arxiv.org/abs/hep-lat/0609068}{{\ttfamily hep-lat/0609068}}].

\bibitem{Aoki:2009sc}
Y.~Aoki, S.~Borsanyi, S.~Durr, Z.~Fodor, S.D.~Katz, S.~Krieg et~al., \emph{{The
  QCD transition temperature: results with physical masses in the continuum
  limit II.}}, \href{https://doi.org/10.1088/1126-6708/2009/06/088}{\emph{JHEP}
  {\bfseries 06} (2009) 088} [\href{https://arxiv.org/abs/0903.4155}{{\ttfamily
  0903.4155}}].

\bibitem{Bazavov:2013yv}
A.~Bazavov and P.~Petreczky, \emph{{Polyakov loop in 2+1 flavor QCD}},
  \href{https://doi.org/10.1103/PhysRevD.87.094505}{\emph{Phys. Rev. D}
  {\bfseries 87} (2013) 094505}
  [\href{https://arxiv.org/abs/1301.3943}{{\ttfamily 1301.3943}}].

\bibitem{Clarke:2019tzf}
D.A.~Clarke, O.~Kaczmarek, F.~Karsch and A.~Lahiri, \emph{{Polyakov Loop
  Susceptibility and Correlators in the Chiral Limit}},
  \href{https://doi.org/10.22323/1.363.0194}{\emph{PoS} {\bfseries LATTICE2019}
  (2020) 194} [\href{https://arxiv.org/abs/1911.07668}{{\ttfamily
  1911.07668}}].

\bibitem{Clarke:2020htu}
D.A.~Clarke, O.~Kaczmarek, F.~Karsch, A.~Lahiri and M.~Sarkar,
  \emph{{Sensitivity of the Polyakov loop and related observables to chiral
  symmetry restoration}},
  \href{https://doi.org/10.1103/PhysRevD.103.L011501}{\emph{Phys. Rev. D}
  {\bfseries 103} (2021) L011501}
  [\href{https://arxiv.org/abs/2008.11678}{{\ttfamily 2008.11678}}].

\bibitem{Clarke:2020clx}
D.A.~Clarke, O.~Kaczmarek, A.~Lahiri and M.~Sarkar, \emph{{Sensitivity of the
  Polyakov loop to chiral symmetry restoration}},
  \href{https://doi.org/10.5506/APHYSPOLBSUPP.14.311}{\emph{Acta Phys. Polon.
  Supp.} {\bfseries 14} (2021) 311}
  [\href{https://arxiv.org/abs/2010.15825}{{\ttfamily 2010.15825}}].

\bibitem{Sarkar:2019jwa}
M.~Sarkar, O.~Kaczmarek, F.~Karsch, A.~Lahiri and C.~Schmidt, \emph{{Conserved
  charge fluctuations with smaller-than-physical quark masses}},
  \href{https://doi.org/10.22323/1.363.0087}{\emph{PoS} {\bfseries LATTICE2019}
  (2019) 087} [\href{https://arxiv.org/abs/1912.11001}{{\ttfamily
  1912.11001}}].

\bibitem{Sarkar:2020soa}
M.~Sarkar, O.~Kaczmarek, F.~Karsch, A.~Lahiri and C.~Schmidt, \emph{{Conserved
  charge fluctuations in the chiral limit}},
  \href{https://doi.org/10.5506/APHYSPOLBSUPP.14.383}{\emph{Acta Phys. Polon.
  Supp.} {\bfseries 14} (2021) 383}
  [\href{https://arxiv.org/abs/2011.00240}{{\ttfamily 2011.00240}}].

\bibitem{Gupta:2015dra}
S.~Gupta and R.~Sharma, \emph{{Lambda Phenomena: the Lambda points of liquid
  Helium and chiral QCD}},
  \href{https://doi.org/10.22323/1.217.0011}{\emph{PoS} {\bfseries CPOD2014}
  (2015) 011} [\href{https://arxiv.org/abs/1503.03206}{{\ttfamily
  1503.03206}}].

\bibitem{HotQCD:2014kol}
{\scshape HotQCD} collaboration, \emph{{Equation of state in ( 2+1 )-flavor
  QCD}}, \href{https://doi.org/10.1103/PhysRevD.90.094503}{\emph{Phys. Rev. D}
  {\bfseries 90} (2014) 094503}
  [\href{https://arxiv.org/abs/1407.6387}{{\ttfamily 1407.6387}}].

\bibitem{Doring:2005ih}
M.~Doring, S.~Ejiri, O.~Kaczmarek, F.~Karsch and E.~Laermann, \emph{{Screening
  of heavy quark free energies at finite temperature and non-zero baryon
  chemical potential}},
  \href{https://doi.org/10.1140/epjc/s2005-02462-y}{\emph{Eur. Phys. J. C}
  {\bfseries 46} (2006) 179}
  [\href{https://arxiv.org/abs/hep-lat/0509001}{{\ttfamily hep-lat/0509001}}].

\bibitem{DElia:2019iis}
M.~D'Elia, F.~Negro, A.~Rucci and F.~Sanfilippo, \emph{{Dependence of the
  static quark free energy on $\mu_B$ and the crossover temperature of $N_f =
  2+1$ QCD}}, \href{https://doi.org/10.1103/PhysRevD.100.054504}{\emph{Phys.
  Rev. D} {\bfseries 100} (2019) 054504}
  [\href{https://arxiv.org/abs/1907.09461}{{\ttfamily 1907.09461}}].

\bibitem{Clarke:2021nrz}
D.A.~Clarke, O.~Kaczmarek, F.~Karsch, A.~Lahiri and M.~Sarkar, \emph{{Imprint
  of chiral symmetry restoration on the Polyakov loop and the heavy quark free
  energy}},  in \emph{{38th International Symposium on Lattice Field Theory}},
  11, 2021 [\href{https://arxiv.org/abs/2111.09844}{{\ttfamily 2111.09844}}].

\bibitem{Bazavov:2016uvm}
A.~Bazavov, N.~Brambilla, H.T.~Ding, P.~Petreczky, H.P.~Schadler, A.~Vairo
  et~al., \emph{{Polyakov loop in 2+1 flavor QCD from low to high
  temperatures}}, \href{https://doi.org/10.1103/PhysRevD.93.114502}{\emph{Phys.
  Rev. D} {\bfseries 93} (2016) 114502}
  [\href{https://arxiv.org/abs/1603.06637}{{\ttfamily 1603.06637}}].

\bibitem{Weber:2016fgn}
{\scshape TUMQCD} collaboration, \emph{{Single quark entropy and the Polyakov
  loop}}, \href{https://doi.org/10.1142/S0217732316300408}{\emph{Mod. Phys.
  Lett. A} {\bfseries 31} (2016) 1630040}
  [\href{https://arxiv.org/abs/1606.06193}{{\ttfamily 1606.06193}}].

\bibitem{Bazavov:2017dus}
A.~Bazavov et~al., \emph{{The QCD Equation of State to $\mathcal{O}(\mu_B^6)$
  from Lattice QCD}},
  \href{https://doi.org/10.1103/PhysRevD.95.054504}{\emph{Phys. Rev. D}
  {\bfseries 95} (2017) 054504}
  [\href{https://arxiv.org/abs/1701.04325}{{\ttfamily 1701.04325}}].

\bibitem{Bonati:2018nut}
C.~Bonati, M.~D'Elia, F.~Negro, F.~Sanfilippo and K.~Zambello, \emph{{Curvature
  of the pseudocritical line in QCD: Taylor expansion matches analytic
  continuation}}, \href{https://doi.org/10.1103/PhysRevD.98.054510}{\emph{Phys.
  Rev. D} {\bfseries 98} (2018) 054510}
  [\href{https://arxiv.org/abs/1805.02960}{{\ttfamily 1805.02960}}].

\bibitem{Bonati:2015bha}
C.~Bonati, M.~D'Elia, M.~Mariti, M.~Mesiti, F.~Negro and F.~Sanfilippo,
  \emph{{Curvature of the chiral pseudocritical line in QCD: Continuum
  extrapolated results}},
  \href{https://doi.org/10.1103/PhysRevD.92.054503}{\emph{Phys. Rev. D}
  {\bfseries 92} (2015) 054503}
  [\href{https://arxiv.org/abs/1507.03571}{{\ttfamily 1507.03571}}].

\bibitem{Kaczmarek:2011zz}
O.~Kaczmarek, F.~Karsch, E.~Laermann, C.~Miao, S.~Mukherjee, P.~Petreczky
  et~al., \emph{{Phase boundary for the chiral transition in (2+1) -flavor QCD
  at small values of the chemical potential}},
  \href{https://doi.org/10.1103/PhysRevD.83.014504}{\emph{Phys. Rev. D}
  {\bfseries 83} (2011) 014504}
  [\href{https://arxiv.org/abs/1011.3130}{{\ttfamily 1011.3130}}].

\bibitem{Sarkar:Lat2021}
M.~Sarkar, ``Critical behavior towards the chiral limit at vanishing and
  non-vanishing chemical potentials - in this conference.''.

\end{thebibliography}\endgroup

\end{document}